\documentclass[twocolumn,floatfix,superscriptaddress,longbibliography,prb]{revtex4-1}

\usepackage{amssymb,amsmath,bm}
\usepackage{graphicx}
\usepackage{epstopdf}
\usepackage{color}
\usepackage{appendix}
\usepackage[colorlinks=true,citecolor=blue,linkcolor=magenta]{hyperref}
\usepackage{float}
\usepackage[normalem]{ulem}

\newcommand{\bbA}{\mathbb{A}}

\newcommand{\bbI}{\mathbb{I}}

\newcommand{\bbM}{\mathbb{M}}

\newcommand{\mC}{\mathcal{C}}

\newcommand{\ket}[1]{\mbox{$| #1 \rangle$}}
\newcommand{\braket}[2]{\mbox{$\langle #1  | #2 \rangle$}}

\definecolor{Green} {rgb}{0.2,0.8,0.2}

\begin{document}

\title{
Variational methods for characterizing matrix product operator symmetries}

\author{Anna Francuz}
\email[corresponding author: ]{anna.francuz@uj.edu.pl}
\affiliation{Jagiellonian University, Institute of Theoretical Physics, {\L}ojasiewicza 11, PL-30348 Krak\'ow, Poland}

\author{Laurens Lootens}
\affiliation{Department of Physics and Astronomy, Ghent University, Krijgslaan 281, S9, B-9000 Ghent, Belgium}

\author{Frank Verstraete}
\affiliation{Department of Physics and Astronomy, Ghent University, Krijgslaan 281, S9, B-9000 Ghent, Belgium}

\author{Jacek Dziarmaga} 
\affiliation{Jagiellonian University, Institute of Theoretical Physics, {\L}ojasiewicza 11, PL-30348 Krak\'ow, Poland}

\date{\today}

\begin{abstract}
We present a method of extracting information about topological order from the ground state of a strongly correlated two-dimensional system represented by an infinite projected entangled pair state (iPEPS). 
As in Phys. Rev. B 101, 041108 (2020) and 102, 235112 (2020) we begin by determining symmetries of the iPEPS represented by infinite matrix product operators (iMPO) that map between the different iPEPS transfer matrix fixed points, to which we apply the fundamental theorem of MPS to find zipper tensors between products of iMPO's that encode fusion properties of the anyons.
The zippers can be combined to extract topological $F$-symbols of the underlying fusion category, which unequivocally identify the topological order of the ground state. We bring the $F$-symbols to the canonical gauge, as well as compute the Drinfeld center of this unitary fusion category to extract the topological $S$ and $T$ matrices encoding mutual- and self-statistics of the emergent anyons.
The algorithm is applied to Abelian toric code, double semion and twisted quantum double of $Z_3$, as well as to non-Abelian double Fibonacci, double Ising, and quantum double of $S_3$ and ${\rm Rep}(S_3)$ string net models.

\end{abstract}

\maketitle


\section{Introduction}

Topologically ordered phases \cite{wen1990topological} support anyonic excitations that open the possibility of realizing fault-tolerant quantum computation \cite{kitaev2003fault-tolerant} by braiding of non-Abelian anyons. Except for a number of exactly solvable models \cite{kitaev2003fault-tolerant, kitaev2006anyons, levin2005string-net}, verifying if a given microscopic Hamiltonian has a topologically ordered ground state was traditionally regarded to be an extremely hard task. Recently, observation of quantized Hall effect in Kitaev-like ruthenium chloride $\alpha$-$RuCl_3$ in magnetic field \cite{KitaevExp} granted the problem with urgent experimental relevance. Intensive experimental search for other Kitaev-like materials is under way\cite{ValentiRecentPRL}.

The density matrix renormalization group (DMRG) \cite{white1992density, white1993density} on a long cylinder used to be the numerical method of choice \cite{yan2011spinliquid, jiang2012identifying, gong2013phase, zhu2013weak, gong2014emergent, zhu2014quantum, gong15global, hu2015topological, zhu15emergence, zhu2015spin, zaletel2016space, zeng2017nature, vaezi2017numerical, zhu2018robust, PollmannKitaevlike, PollmannKitaev}. In the limit of infinitely long cylinders, DMRG naturally produces ground states with well-defined anyonic flux from which one can obtain characterization of a topological order via so-called topological $S$ and $T$ matrices \cite{cincio2013characterizing}. Since the proposal of Ref. \onlinecite{cincio2013characterizing}, this approach has become a common practice \cite{he2014chiral, zhu2014topological, zhu2014chiral, bauer2014chiral, zhu2015fractional, grushin15characterization, he2015kagome, he15distinct, he15bosonic,  geraedts15competing, mong2015fibonacci, he17realizing, stoudenmire2015assembling, he17signatures, Saadatmand20016symmetry, hickey2016haldane, zaletel2017measuring, zeng2018tunning}.

Unfortunately, the cost of a DMRG simulation grows exponentially with the circumference of cylinder, limiting this approach to thin cylinders (up to a width of $\simeq 14$ sites) and thus to short correlation lengths (up to $1-2$ sites). Instead, infinite projected entangled pair states (iPEPS) in principle allow for much longer correlation lengths \cite{verstraete2004renormalization, murg2007variational, verstraete2008matrix}. A unique ground state on an infinite lattice can be represented by an iPEPS that is either a variational ansatz \cite{chiralpeps} or a result of numerical optimization \cite{varCorboz,topoAF1,chiralPoilblanc}. Either way it seems feasible to obtain an iPEPS with non-Abelian chiral topological order \cite{chiralPoilblanc,chiralpeps}. The ferromagnetic Kitaev model in a weak $(1,1,1)$ magnetic field supports non-Abelian chiral topological order \cite{kitaev2006anyons,PollmannKitaev} and Ref. \onlinecite{KitaevExp} provides the first experimental realization of this universality class. However, as the magnetic field is a tiny perturbation of a critical state, the correlation length should be large driving the problem beyond accurate DMRG simulation on a thin cylinder.

When wrapped on a cylinder the iPEPS becomes a superposition of degenerate ground states with definite anyonic fluxes. In the realm of the string-net models it is possible \cite{topoAF1,topoAF2} to produce a PEPS-like tensor network for each ground state with well-defined flux. Such tensor networks are suitable for extracting topological $S$ and $T$ matrices by computing overlaps between the ground states. Furthermore, they allow for computation of topological second Renyi entropy directly in the limit of infinite cylinder's width. The approach of Refs. \onlinecite{topoAF1,topoAF2} does not assume clean realization of certain symmetries on the bond indices, in contrast to \cite{burak2014characterizing, bultinck2017anyons, iqbal2018study, fernandez2016constructing}. This has been demonstrated in Ref. \onlinecite{topoAF1} by examples of toric code and double semions perturbed away from a fixed point towards a ferromagnetic phase as well as for the numerical iPEPS representing the ground state of the Kitaev model in the gapped phase. The same approch was generalized to non-Abelian topological order in Ref. \onlinecite{topoAF2}. The method does not require restoring the symmetries by suitable gauge transformations of a numerical iPEPS, a feat that was accomplished in Ref. \onlinecite{topoCorboz} for the toric code with a perturbation. It is also not necessary to optimize symmetry-constrained iPEPS tensors as in Ref. \onlinecite{iqbal2020order}. Finally, it also has much lower numerical cost than methods based on the tensor renormalization group \cite{he2014modular}.

In this work we reconsider the string-net models. Similarly as in Refs. \onlinecite{topoAF1,topoAF2}, for a given iPEPS we obtain numerically its infinite matrix product operator (iMPO) symmetries. Products of the iMPO-symmetries realize fusion rules of the corresponding anyons of a unitary fusion category (UFC) $\mC$. We use the fundamental theorem of \textit{matrix product states} (MPS) \cite{fundtheorem,IrreducibleMPS} and apply it to the iMPO products in order to classify topological order through its related fusion categories. The fundamental theorem of MPS has already been widely used in characterization of phases of both 1D and 2D gapped systems \cite{WenClasstfyingPhases,SchuchClassifyingPhases} as well as the construction of exact renormalization fixed point representations of string-nets with iPEPS\cite{bultinck2017anyons,burak2014characterizing}. The theorem allows us to construct gauge transformations (zippers) between products of iMPO's and their fusion outcomes. The zippers encode information on fusion properties of the corresponding anyons, and they can be combined in order to extract the $F$-symbols of the underlying UFC $\mC$ describing the topological order unequivocally. The different ground states and possible anyonic excitations of the string-net model are actually described by the Drinfeld center $Z(\mC)$, and different UFCs $\mC$ associated to the iMPO symmetries can give the same topological order if their centers are isomorphic \cite{Lootens_2021}. To deal with this redundancy, we compute the center by constructing idempotents of the tube algebra and compute invariants such as the topological $S$ and $T$ matrices which encode mutual- and self-statistics of the emergent anyons. While the $S$ and $T$ matrices provide a useful characterization of the type of topological order, in general they do not uniquely specify the modular category $Z(\mC)$ \cite{mignard2021modular}. By explicitly constructing $Z(\mC)$, our approach does not suffer from this problem.

The method we use has similarities with previous approaches where one looks for string-like operators on the physical level that commute with the Hamiltonian called ribbon operators \cite{bridgeman2016detecting}. An important fact is that in these approaches, when moving away from the fixed point, these ribbon operators get dressed \cite{bravyi2010topological} and their width is proportional to the correlation length. In contrast, in our approach, the MPO symmetries are not fattened when perturbing the system away from the fixed point since they act purely on the virtual level.

The paper is organized in sections \ref{sec:Z}...\ref{sec:FtoST} where we gradually introduce subsequent elements of the algorithm. Most sections open with a general part introducing a new concept. Then a series of subsections follows illustrating the general concept with a series of examples: toric code and double semions, Fibonacci, twisted quantum double of $Z_3$, Ising string net, and quantum double of $S_3$ and ${\rm Rep}(S_3)$. In the end the algorithm is summarized in section \ref{sec:summary}. A detailed plan is as follows.

In Sec. \ref{sec:Z} we define fixed points of the iPEPS transfer matrix in the form of iMPS and introduce iMPO symmetries that map between different fixed points. We also identify fusion rules of the iMPO symmetries that are isomorphic with the fusion rules of some input category $\mC$.
In Sec. \ref{sec:X} we introduce $X$ zippers that are gauge transformations between products of two iMPO symmetries acting on a trivial fixed point of the transfer matrix and a single iMPO symmetry applied to the same trivial fixed point. We distinguish between up and down $X$ zippers for, respectively, up and down fixed points.
In Sec. \ref{sec:Y} we introduce and construct more elementary $Y$ zippers. Each $Y$ zipper is a gauge transformation between a product of an iMPO symmetry and a fixed point of the transfer matrix and the resulting fixed point. $X$ zippers can be constructed out of the elementary $Y$ zippers.
In Sec. \ref{sec:normX} pairs of complementary left and right $X$ zippers are normalized to become pairs of gauge and inverse gauge transformations. In particular, non-trivial normalization between up and down zippers is imposed.
In Sec. \ref{sec:F} we construct $F$ symbols out of the normalized up and down $X$ zippers. The fusion symbols have arbitrary/random numerical gauge.
In Sec. \ref{sec:Fcanonical} we parameterize the gauge freedom and outline how the $F$ symbols can be brought to textbook canonical gauge that allows to identify the topological order.
In Sec. \ref{sec:FtoST} we algebraically construct the gauge-invariant central idempotents of the tube algebra made of the zippers, which when inserted into iPEPS can be thought of as projectors onto minimally entangled states (MES). However, here we do not construct the MES but use the central idempotents together with the tube algebra to directly extract topological $S$ and $T$ matrices. Unlike the $F$ symbols, the $S$ and $T$ matrices are gauge-invariant observables with a physical interpretation of statistics of the emergent anyons. In contrast to Ref. \onlinecite{topoAF1,topoAF2}, here they are obtained by algebraic manipulation from the $F$-symbols, the calculation of which is a purely 1D problem, which significantly reduces the complexity of the numerical algorithms whereas the calculations of $S$ and $T$ matrices via overlaps \cite{topoAF1,topoAF2} between different MES is done on an infinite 2D lattice. Thus the route via the $F$-symbols is an alternative that is potentially more stable numerically.
The paper is closed with a brief summary of the algorithm in section \ref{sec:summary} and an outlook towards future applications.

\begin{figure}[t!]
\includegraphics[width=\columnwidth]{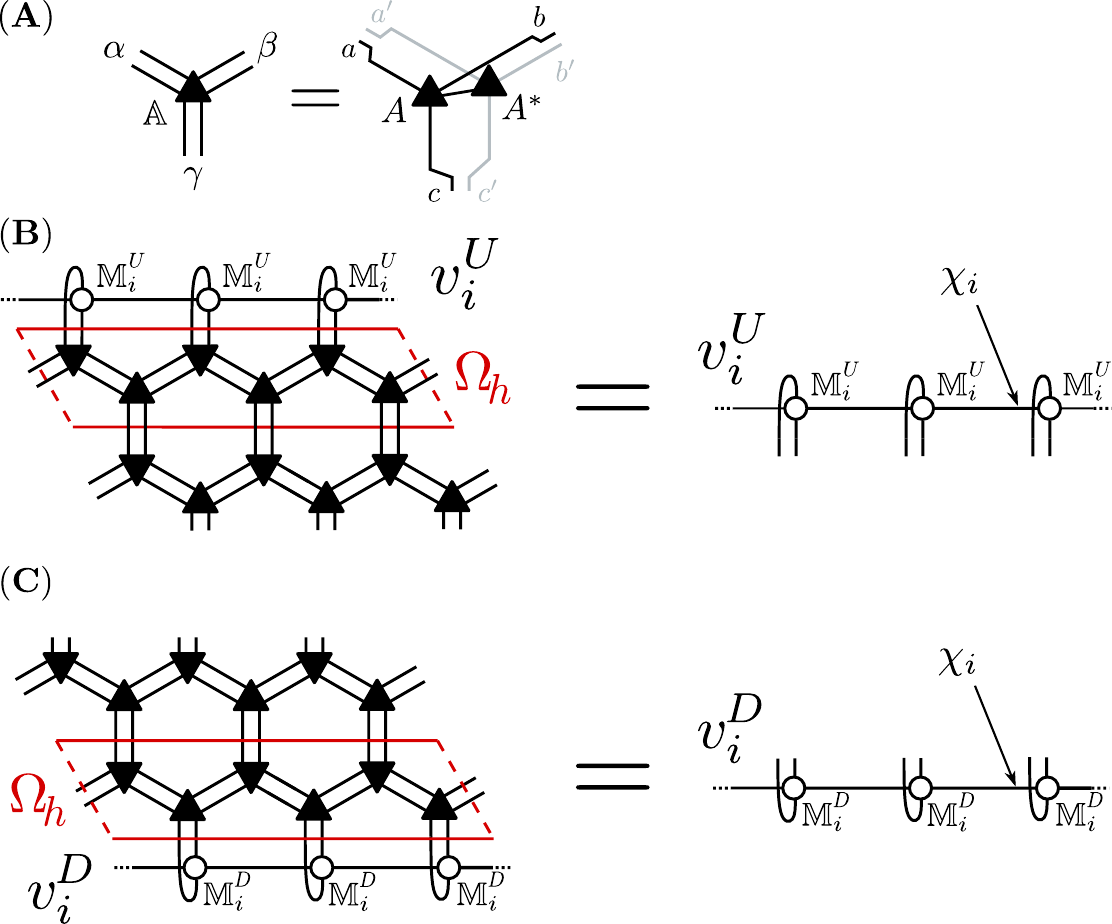}
\caption{
{\bf Transfer matrix.}
In (A), 
graphical representation of a double iPEPS tensor $\bbA$ that is made out of an iPEPS tensor $A$ contracted through a physical index with its complex conjugate $A^*$.
In (B) and (C), 
a horizontal row of $\bbA$ makes a horizontal transfer matrix $\Omega_h$. Its leading up-eigenvectors, $\left(v_i^U\right|$ and down-eigenvectors, $\left(v_i^D\right|$, with the leading degenerate eigenvalue $1$, can be obtained with the VUMPS algorithm \cite{VUMPS2,VUMPS}. The eigenvector can be reshaped into an iMPO form $v_i^U$. The uniform $v_i^U$ is made of tensors $\bbM_i^{U}$ with bond dimension $\chi_i$.
}
\label{fig:TM}
\end{figure}

\section{Numerical symmetries and fusion algebra}
\label{sec:Z}

The iPEPS representing the ground state on an infinite lattice, $\ket{\psi}$, is assumed to be normalized: $\braket{\psi}{\psi}=n$. Its norm, which is a contraction between the iPEPS (ket) and its complex conjugate (bra), is a 2D tensor network made of double iPEPS tensors shown in Fig. \ref{fig:TM}(A). Each row of the network is a horizontal transfer matrix $\Omega_h$ in Fig. \ref{fig:TM}(B). The transfer matrix has several leading up-eigenvectors, $\left.\vert v_i^U\right)$ numbered by $i$, whose degenerate leading eigenvalue is $1$ (hence the double iPEPS with $n$ leading eigenvectors is normalized to $n$). These {\it boundary fixed points} can be reshaped as iMPO's, $v_i^U$, acting between virtual bra and ket indices. Together with their corresponding biorthonormal down-eigenvectors, $\left(v_i^D\vert\right.$, that can be also reshaped as iMPO, $v_i^D$, they satisfy:
\begin{eqnarray}
\Omega_h &\approx& 1 \sum_{i=1}^n \left.\vert v_i^U\right)\left(v_i^D\vert\right.,  \label{eq:eigvecs1} \\
\delta_{ij} &=& \left(v_i^U\vert v_j^D\right) = {\rm Tr}~ \left( v_i^U \right)^T v_j^D.  \label{eq:eigvecs2}
\end{eqnarray}

The first and the most important step to identify the topological order is finding the virtual iMPO symmetries of the iPEPS as their existence is a necessary condition for the iPEPS to exhibit topological order. As described in Ref. \onlinecite{topoAF1,topoAF2}, the iMPO symmetries $Z_a$ are found numerically as operators acting between different iMPO boundary fixed points $v_i$:
\begin{eqnarray}
&& v_i^U \cdot Z_a   = \sum_k \delta_{iak} v_k^U, \nonumber\\
&& v_i^D \cdot Z_a^T = \sum_k \bar{\delta}_{iak} v_k^D.
\label{eq:mpo_sym1}
\end{eqnarray}
Here $\delta_{iak}$ and $\bar{\delta}_{iak}$ take values either $0$ or $1$ and in general they do not have to be the same. A trivial $v_1^{U,D}$ can be identified such that its trace with all the iMPO symmetries is equal 1: $\mathrm{Tr}(v^U_1\cdot Z_a\cdot v^D_1 \cdot Z_a^\dagger) = 1$. In particular for the up-eigenvector $v^U_1$ all other $v^U_{i>1}$ are obtained from it by the action of corresponding iMPO symmetries:
\begin{eqnarray}
v_1^U \cdot Z_a   = v_a^U,
\label{eq:mpo_sym1}
\end{eqnarray}
while at the same time for the down- eigenvectors:
\begin{eqnarray}
v_1^D \cdot Z_a^T = v_{\bar{a}}^D.
\label{eq:mpo_sym3}
\end{eqnarray}
Each symmetry $Z_a$, including the trivial $Z_1 = \mathbb{I}$, represents certain anyon type $a$. The symmetries form a representation of a fusion ring,
\begin{eqnarray}
Z_a\cdot Z_b = \sum_c N^c_{ab} Z_c,
\label{eq:mpo_sym2}
\end{eqnarray}
where $N^c_{ab}$ take values either $0$ or $1$. For an Abelian ring $N^c_{ab} = N^c_{ba}$, but in general the order of multiplications of iMPO symmetries has to be strictly controlled. This way, with just a little more numerical effort than required to obtain the iPEPS itself, the anyonic fusion ring can be identified as: $a\cdot b = \sum_c N^c_{ab} c$. 

Indeed, the iMPO symmetries can be obtained by variational minimization of a convenient quadratic cost function with respect to tensor $z_a$ of the uniform iMPO $Z_a$:
\begin{equation}
\vert v_1^U\cdot Z_{a} - v_a^U\vert^2 + \vert v^D_1\cdot Z_a^T - v^D_a \vert^2,
\label{eq:Zmin}
\end{equation}
when the action of $Z_a$ is equal to the action of its transpose on the support subspace of the boundary eigenvectors and otherwise:
\begin{equation}
\vert v_1^U\cdot Z_{a} - v_a^U\vert^2 + \vert v^D_a\cdot Z_a^T - v^D_1 \vert^2,
\label{eq:Zmin2}
\end{equation}

In order to minimize the effect of the unnecessary modes in the nullspace of an iMPO symmetry acting on the up and down boundary eigenvectors the bond dimension of $Z_a$, $\chi_a$, has to be the minimal one that still allows the cost function to be nullified. By definition, this cost function guarantees the correct action of the symmetries on the boundary fixed points but not the ``abstract'' fusion ring \eqref{eq:mpo_sym2}. However, the algebra is satisfied in a weaker sense: 
\begin{eqnarray}
&& v_i^U Z_a\cdot Z_b = \sum_c N^c_{ab}~ v_i^U Z_c, \\
&& v_i^D Z_a^T\cdot Z_b^T = \sum_d N^d_{ba}~ v_i^D Z_d^T,
\label{eq:vZZ}
\end{eqnarray}
i.e., when applied to any boundary fixed point. This is all that we need in the following construction.

We completed these numerical procedures in the following models. Some of the examples are the same as in Ref. \onlinecite{topoAF2} but notice that here the cost function \eqref{eq:Zmin} is more demanding because it has two terms instead of just one as in Ref. \onlinecite{topoAF2}. Numerically there is a freedom of the global phase of the eigenvectors $v_i^{U,D}$, which can be partially eliminated (up to minus sign) by requiring their Hermiticity (when applicable). In general the random global phases change the fusion rules, so that only their absolute values are 0 or 1, $\vert N^c_{ab}\vert$ = 0,1. However in all the examples below the random global phases are adjusted so that all $N^c_{ab}$ are real, either 0 or 1.

\subsection{Toric code and double semions}

For analytic fixed point tensors defined in appendix \ref{app:tensors} transfer matrix $\Omega_h$ has $2$ numerical boundary fixed points $v^{U,D}_{1,2}$ and one non-trivial numerical iMPO symmetry $Z_2$ which fulfills the $Z_2$ algebra:
\begin{eqnarray}
\left.\begin{array}{l} 
v_1^U\cdot Z_2 = v_2^U \\
v_2^U\cdot Z_2 = v_1^U 
\end{array} \right\rbrace ~\Rightarrow~ 
Z_2\cdot Z_2 = \mathbb{I}.
\label{Ztc}
\end{eqnarray}
The cost function \eqref{eq:Zmin} was minimized to zero within machine precision. The fusion rules can be summarized as
\begin{equation}
    N^1_{11} = N^1_{22} = 1
\end{equation}
with all possible permutation of indices.
It has to be strongly emphasized that in general the numerical $Z_2$ iMPO symmetry is not neccesarily nullified outside of the support subspace of the boundary eigenvectors, therefore the ring on the right of (\ref{Ztc}) is valid only in the sense of the equalities on the left. The same reservation applies to all fusion rules to be identified numerically in the rest of this paper.

\subsection{Twisted quantum double of $Z_3$}

The transfer matrix $\Omega_h$ has 3 boundary fixed points $v_{1,2,3}^{U,D}$, out of which only one, $v_1^U$ and corresponding $v_1^D$, is Hermitian and it plays the role of the trivial boundary. The other two  boundary fixed points are their own Hermitian conjugates : $v_2^{U,D} = (v_3^{U,D})^\dagger $. Here we notice that the quadratic form to minimize for the iMPO symmetry (with bond dimension $\chi=2$) is Eq. \eqref{eq:Zmin2}, which one realizes after finding the symmetry from the single condition $|v_1^U\cdot Z_a - v_a^U|^2$ and checking its algebra:
\begin{eqnarray}
\left.\begin{array}{l} 
v_1^U\cdot Z_q = v_2^U \\
v_2^U\cdot Z_q = v_3^U \\ 
v_3^U\cdot Z_q = v_1^U \end{array}, ~
\begin{array}{l} 
v_1^U\cdot Z_{q^*} = v_3^U \\
v_2^U\cdot Z_{q^*} = v_1^U \\ 
v_3^U\cdot Z_{q^*} = v_2^U \end{array}
\right\rbrace &\Rightarrow& 
Z_q Z_{q^*} = \mathbb{I}, \\
\left.\begin{array}{l} 
v_1^D\cdot Z^T_q = v_3^D \\
v_2^D\cdot Z^T_q = v_1^D \\ 
v_3^D\cdot Z^T_q = v_2^D \end{array},~
\begin{array}{l} 
v_1^D\cdot Z^T_{q^*} = v_2^D \\
v_2^D\cdot Z^T_{q^*} = v_3^D \\ 
v_3^D\cdot Z^T_{q^*} = v_1^D \end{array}
\right\rbrace &\Rightarrow& 
Z_{q^*} Z_q= \mathbb{I}.
\end{eqnarray}
The two iMPO symmetries $Z_q, Z_{q^*}$, are denoted with the subscripts $q,q^* = \mathrm{e}^{\pm2i\pi/3}$. Despite different fusions with the eigenvectors $\delta_{iak} \neq \bar{\delta}_{iak}$ the fusion rules of the iMPO symmetries are given by the following non-zero elements of the fusion tensor
\begin{eqnarray}
&\forall_{i=1,q,q^*}~  N^i_{1i}= N^i_{i1} = 1,& \nonumber \\
&N^1_{qq^*}= N^1_{q^*q}= N^{q^*}_{qq}= N^q_{q^*q^*} = 1.&
\end{eqnarray}
In this case the anyon types $q$ and $q^*$ are the inverses of each other, which justifies the labeling.

\subsection{Fibonacci string-net}

Here we employed the iPEPS tensors for a fixed point Fibonacci string net model presented in appendix \ref{app:tensors}. The transfer matrix $\Omega_h$ has $2$ numerical boundary fixed points, $v^{U,D}_{1,2}$, and one non-trivial numerical iMPO symmetry $Z_\tau$ which fulfills:
\begin{eqnarray}
\left.\begin{array}{l} v_1^U\cdot Z_\tau = v^U_2 \\
v^U_2\cdot Z_\tau = v^U_1 + v^U_2 \end{array} \right\rbrace ~\Rightarrow~ Z_\tau\cdot Z_\tau = \mathbb{I}+Z_\tau.
\label{ZF}
\end{eqnarray}
Again, the cost function \eqref{eq:Zmin} was minimized to vanish up to machine precision and the fusion on the right holds only in the sense of the equalities on the left. The fusion algebra on the right of \eqref{ZF} allows us to label the iMPO symmetry with a non-Abelian Fibonacci anyon $\tau$. The fusion rules can be summarized as
\begin{equation}
    N^1_{11} = N^1_{\tau\tau} = N^\tau_{\tau\tau} = 1.
\end{equation}
with all possible permutation of indices.

\subsection{Fibonacci string-net with local filtering}

In order to drive the iPEPS away from a fixed point and introduce a finite correlation length we apply the local filtering \cite{per1,per2,per3} to the fixed point of the Fibonacci string-net model. The modification has the following form:
\begin{equation}
    \vert \Psi\rangle ~\rightarrow~ \prod_i \mathrm{e}^{\beta\sigma^z_i}\vert \Psi\rangle, 
    \label{eq:beta}
\end{equation}
where $i$ runs over all physical indices, $\sigma^z$ is the Pauli matrix, and $\beta$ is a parameter. Correlation lengths $\xi$ are listed in table \ref{table:tab1}. In the table we also present errors of the two terms appearing in the cost function \eqref{eq:Zmin}: $\epsilon^U_Z = \vert(1-\left(v_1^U\cdot Z_\tau\vert v_2^U\right)\vert$ and $\epsilon^D_Z = \vert 1-\left(v_1^D\cdot Z^T_\tau \vert v_2^D\right)\vert $. The difference between errors of $\epsilon^U_Z$, $\epsilon^D_Z$ arises from the fact that with growing correlation length it becomes harder to nullify both errors at the same time, therefore, in order to ensure convergence, in the step where we find an optimal update as described in Ref. \onlinecite{varCorboz}, we use only one of the conditions to be best approximated, namely ${\epsilon^U_Z = \vert v_1^U\cdot Z_\tau - v_2^U\vert^2}$.

\begin{table}[H]
\centering \begin{tabular}{c|c|c|c}
$\beta$ & $\xi$ &      $\epsilon_Z^U$     & $\epsilon_Z^D$          \\ \hline
0.01    & 0.23  &  $\mathcal{O}(10^{-7})$ & $\mathcal{O}(10^{-7})$  \\
0.05    & 0.42  &  $\mathcal{O}(10^{-4})$ & $\mathcal{O}(10^{-3})$  \\
0.12    & 1.04  &  $\mathcal{O}(10^{-4})$ & $\mathcal{O}(10^{-2})$  \\
0.15    & 2.32  &  $\mathcal{O}(10^{-2})$ & 0.05 
\end{tabular}
\caption{The errors $\epsilon^U_Z$, $\epsilon^D_Z$ of the fusion ring of the the numerical iMPO symmetry with the corresponding up- and down- eigenvectors $v_1^{U,D}$. Parameter $\beta$ represents the perturbation strength from Eq. \eqref{eq:beta}, while $\xi$ is the corresponding correlation length calculated from the second leading eigenvalue of the iPEPS transfer matrix in the environment of the boundary eigenvectors.}
\label{table:tab1}
\end{table}

\subsection{Ising string net}

Here again we employed the iPEPS tensors for a fixed point Ising string net model presented in appendix \ref{app:tensors}. This time each TM has $3$ numerical boundary fixed points, $v^{U,D}_{1,2,3}$, corresponding to $3$ anyon types of the input category: $1,\sigma,\psi$. We found two non-trivial iMPO symmetries, labelled as $Z_\sigma$ and $Z_\psi$. The fixed points and the symmetries are related by the following set of equations:
\begin{eqnarray}
\left.\begin{array}{l} 
v_1^U\cdot Z_\psi = v_2^U \\
v_2^U\cdot Z_\psi = v_1^U \\ 
v_3^U\cdot Z_\psi = v_3^U \end{array} \right\rbrace &\Rightarrow& 
Z_\psi\cdot Z_\psi = \mathbb{I}, \\
\left.\begin{array}{l}
v_1^U\cdot Z_\sigma = v_3^U \\
v_2^U\cdot Z_\sigma = v_3^U \\ 
v_3^U\cdot Z_\sigma = v_1^U+v_2^U \end{array} \right\rbrace &\Rightarrow& 
Z_\sigma\cdot Z_\sigma = \mathbb{I}+Z_\psi.
\end{eqnarray}
The cost function \eqref{eq:Zmin} was minimized to numerical zero. Furthermore, we verified that with machine precision the symmetries satisfy:
\begin{equation}
    v_i^U Z_\sigma\cdot Z_\psi = v_i^U Z_\sigma ~\Rightarrow~  Z_\sigma\cdot Z_\psi = Z_\sigma
\end{equation}
The equations justify labelling of the symmetries. The fusion rules can be summarized as
\begin{equation}
    N^1_{11} = N^1_{\sigma\sigma} = N^1_{\psi\psi} = N^\sigma_{\psi\psi} = 1
\end{equation}
with all possible permutation of indices.

\subsection{Quantum double of $S_3$ and $\text{Rep}(S_3)$ string-net}
In this section we analyze two different iPEPS representations from Ref. \onlinecite{Lootens_2021} for the quantum double $S_3$ and the $\text{Rep}(S_3)$ string-net model, with MPO symmetries respectively given by UFCs $\mC_1 = \text{Rep}(S_3)$ and $\mC_2 = \text{Vec}_{S_3}$. These two iPEPS representations describe the same topologically ordered phase since $Z(\text{Rep}(S_3)) = Z(\text{Vec}_{S_3})$.

\subsubsection*{$\text{Rep}(S_3)$ MPO symmetries}
In this representation iPEPS tensor has virtual bond dimension $D=6$ and its related transfer matrix $\Omega_h$ has 3 leading eigenvectors $v^{U,D}_{1,2,3}$ corresponding to 3 anyon types $1,\pi,\psi$. There are two nontrivial iMPO symmetries, with corresponding labels $\pi,\psi$ and they fulfill the following fusion rules with the eigenvectors:
\begin{align}
\left.\begin{array}{l} 
v_1^U\cdot Z_\psi = v_3^U \\
v_2^U\cdot Z_\psi = v_2^U + v_3^U \\ 
v_3^U\cdot Z_\psi = v_1^U \end{array} \right\rbrace &\Rightarrow
Z_\psi\cdot Z_\psi = \mathbb{I}, \\
\left.\begin{array}{l} 
v_1^U\cdot Z_\pi = v_3^U \\
v_2^U\cdot Z_\pi = v_1^U + v_2^U + v_3^U \\ 
v_3^U\cdot Z_\pi = v_1^U \end{array} \right\rbrace &\Rightarrow 
Z_\sigma\cdot Z_\sigma = \mathbb{I}+Z_\psi+Z_\pi.
\end{align}
The same set of equations can be written for the down-eigenvectors. Moreover we observe that:
\begin{equation}
    v_i^U Z_\pi\cdot Z_\psi = v_i^U Z_\pi \Rightarrow Z_\pi\cdot Z_\psi = Z_\pi,
\end{equation}
which enables identification of all allowed fusion rules:
\begin{equation}
    N^1_{11} = N^1_{\pi\pi} = N^1_{\psi\psi} = N^\sigma_{\psi\psi}  = N^\pi_{\pi\pi}= 1
\end{equation}
with all possible permutation of indices.

\subsubsection*{$\text{Vec}_{S_3}$ MPO symmetries}

In this representation the iPEPS tensor has bond dimension that is just $D=4$ while its related transfer matrix $\Omega_h$ has degeneracy 6 corresponding to 6 leading eigenvectors $v^{U,D}_{1,2,3,4,5,6}$. There are 5 nontrivial iMPO symmetries $Z_a$, which are all product iMPOs. There is only one eigenvector, which we label as identity, for which $\text{Tr}(v_1^U\cdot Z_a\cdot v_1^D\cdot (Z_a)^\dagger) = 1$ for all $a = 1,...,6$. All the remaining up and down eigenvectors can be obtained from $v_1^{U,D}$ by proper action of the iMPO symmetries:
\begin{eqnarray}
   && v_1^U\cdot Z_a = v_a^U,~ \forall a = 1,2,3,4,5,6 \nonumber \\
   && v_1^D\cdot Z^T_a = v_a^D, ~\forall a = 1,2,3,4, \nonumber \\ 
   && v_1^D\cdot Z^T_5 = v_6^D, ~v_1^D\cdot Z^T_6 = v_5^D
\end{eqnarray}
In this case the fusion ring is non-Abelian in the sense that $N^c_{ab} \neq N^c_{ba}$ and apart from trivial fusion rules $N^a_{1a} = N^a_{a1} = 1$, there are 25 non trivial ones, all equal 1:
\begin{eqnarray}
   && N^1_{22}, N^1_{33}, N^1_{44}, N^1_{65}, N^1_{56}, N^2_{63}, N^2_{54}, N^2_{35}, N^2_{46}, \nonumber \\
   && N^3_{52}, N^3_{64}, N^3_{45}, N^3_{26}, N^4_{62}, N^4_{53}, N^4_{25}, N^4_{36}, \nonumber \\
   && N^5_{32}, N^5_{43}, N^5_{24}, N^5_{66}, N^6_{42}, N^6_{23}, N^6_{34}, N^6_{55}.
\end{eqnarray}
From this we notice that iMPO symmetries $Z_a$ for $a = 1,2,3,4$ are self-inverse, while $Z_5$ is the inverse of $Z_6$.

\section{Numerical $X$ zippers}
\label{sec:X}

In this work we employ the fundamental theorem of MPS\cite{fundtheorem,IrreducibleMPS} according to which there exist an invertible gauge transformation $G_l$ between two tensors $A^i$ and $B^i$, where $i$ is the ``physical'' index, 
both in a canonical form, generating equal iMPS's such that 
\begin{equation}
    A^i = G_l \cdot B^i \cdot G_r,
\end{equation}
where $G_r=G_l^{-1}$. It can be further extended to iMPO and products of iMPO's where, e.g., $A=v_i\cdot Z_a$ and $B = v_k$. In that case the bond dimension of the product $v_i\cdot Z_a$ is usually bigger than the bond dimension of $v_k$: $\chi_i\cdot \chi_z > \chi_k$. Therefore, the gauge transformation $G_l$ is actually a composition of an isometry $U$ of dimensions $(\chi_i\cdot \chi_z, \chi_k)$ and an invertible $\chi_k\times\chi_k$ matrix $g$. $G_r$ is a pseudo-inverse of $G_l$ and vice versa.

Due to the algebra of iMPO symmetries \eqref{eq:mpo_sym2}, which is fulfilled only when acting on the boundary fixed points \eqref{eq:vZZ}, our goal is to find zipper tensors $X^c_{ab}$ which serve as gauge transformations between products $v_1\cdot Z_a\cdot Z_b$ and $v_1\cdot Z_c$:
\begin{equation}
    v_1\cdot Z_a\cdot Z_b ~~\stackrel{X^c_{ab}}{\longleftrightarrow}~~ v_1\cdot Z_c
\end{equation}
Here we consider only the trivial fixed point $v_1$ to make sure the fusions actually occur only between the iMPO symmetries $Z$. 

\begin{figure}[H]
\centering\includegraphics[scale=0.4]{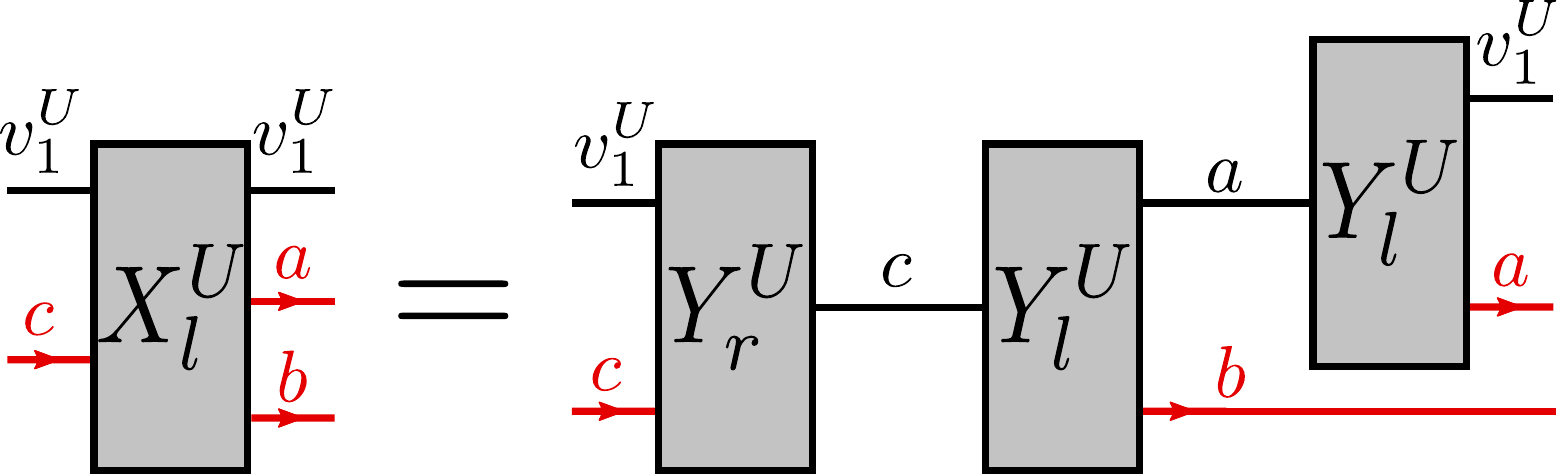}
\caption{Contraction of $Y$ zippers that makes an $X$ zipper. 
Here we show only $U$-zippers but similar equations hold for their $D$ counterparts. We distinguish between left, $l$, and right, $r$, zippers. Contraction of a left zipper with its corresponding right zipper yields an identity. One is a pseudo-inverse of the other.
}
\label{fig:Xzip}
\end{figure}

This goal can be achieved in two steps, first by obtaining smaller zippers $Y^k_{ia}$ which fuse a product $v_i\cdot Z_a$ into a single MPO $v_k$. The second step is the proper contraction of zippers $Y^k_{ia}$ to form $X^c_{ab}$ as shown in Fig. \ref{fig:Xzip}.

\section{Numerical $Y$ zippers}
\label{sec:Y}

The iMPO symmetry algebra, $Z_a$, includes a trivial symmetry $Z_1=\mathbb{I}$ corresponding to the trivial anyon type. 
This identity iMPO is a product of identity matrices and has bond dimension $\chi_1=1$. Therefore, all the zippers 
$(Y_l)^j_{j1}$, together with their $r$ (inverse) counterparts, are trivial identity matrices of dimensions $\chi_j\times \chi_j$.

From now on we focus the attention on non-trivial zippers between the left and right hand side of the equation $v_j \cdot Z_a = \sum_k \delta_{jak}v_k$ with $a>1$. The product MPO tensor $M=v_j\cdot Z_a$ is either normal, for which a transfer matrix of ${\rm Tr} MM^\dagger$ has only one leading eigenvalue equal $1$, or a direct sum of normal tensors, so the transfer matrix has several degenerate leading eigenvalues equal $1$. In both cases we proceed by bringing the tensors $M$ into left-canonical form using a repeated QR decomposition. For a fixed point of QR decomposition the relation between the initial tensor and the converged canonical form is:
\begin{equation}
L\cdot M^i = M_L^i \cdot L,
\end{equation}
which means that the transformation bringing the tensor $M^i$ into its canonical form $M_L^i$ is:
\begin{equation}
L\cdot M^i\cdot {\rm pinv}(L) = M_L^i.
\end{equation}
Here $i$ denotes ``physical'' indices of tensor $M$. The pseudo-inverse deals with singularity due to too large bond dimension of $M$: $\chi_j\cdot\chi_a>\sum_k\chi_k$. 

In the next step we reduce the bond dimension for $M^i_L$ and find the gauge transformation relating it with one of the $v_k^i$ tensors. Towards this end we construct a mixed transfer matrix for ${\rm Tr} M_L v_k^\dag$. Its left fixed point, $\sigma_L$, is an isometry of dimension $\chi_j\cdot\chi_a\times \chi_k$ truncating the left-canonized product $M_L$ to $v_k$: 
\begin{equation}
\sigma_L^T\cdot M_L^i\cdot {\rm pinv }(\sigma_L^T) = v_k^i.   
\end{equation}
Putting the isometry together with the gauge transformation $L$ we can write:
\begin{equation}
(Y_l)^k_{ja} \left(v_j\cdot Z_a\right)^i (Y_r)^k_{ja} = v_k^i, 
\label{YMY}
\end{equation}
where
\begin{eqnarray}
(Y_l)^k_{ja} &=& \sigma_L^T\cdot L,~ \\
(Y_r)^k_{ja} &=&   {\rm pinv}(L)\cdot {\rm pinv}(\sigma_L^T).
\end{eqnarray}
In diagramatic form equation \eqref{YMY} is:
\begin{figure}[H]
\centering\includegraphics[scale=0.4]{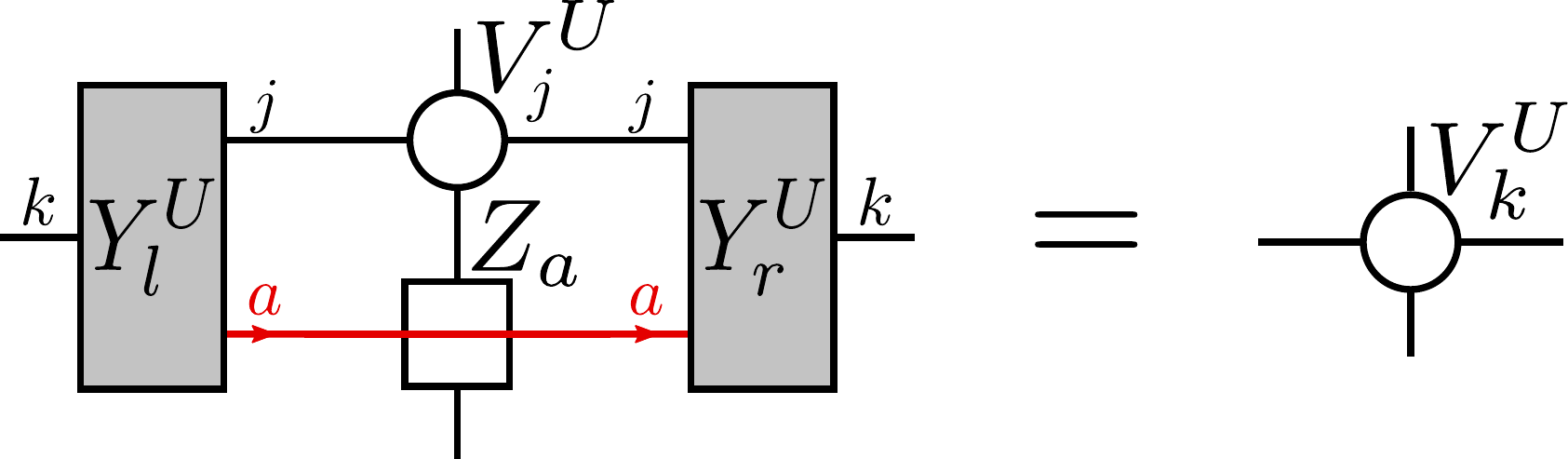}
\caption{A pair of $Y$ zippers, $(Y_l^U)^k_{ja}$ and $(Y_r^U)^k_{ja}$, contracts a product of $v_j^U\cdot Z_a$ into $v_k^U$.}
\label{fig:Yzip}
\end{figure}

In all equations above we did not include the labels of the eigenvectors $v_i$, as there are two sets of them: $v_i^U$ and $v_i^D$. However, the procedure is the same for both sets with a sole difference that for the down eigenvectors  we need to use the transpose $Z_a^T$ in place of $Z_a$.  

\section{Normalization of $X$ zippers}
\label{sec:normX}

\begin{figure}[t!]
\centering\includegraphics[scale=0.25]{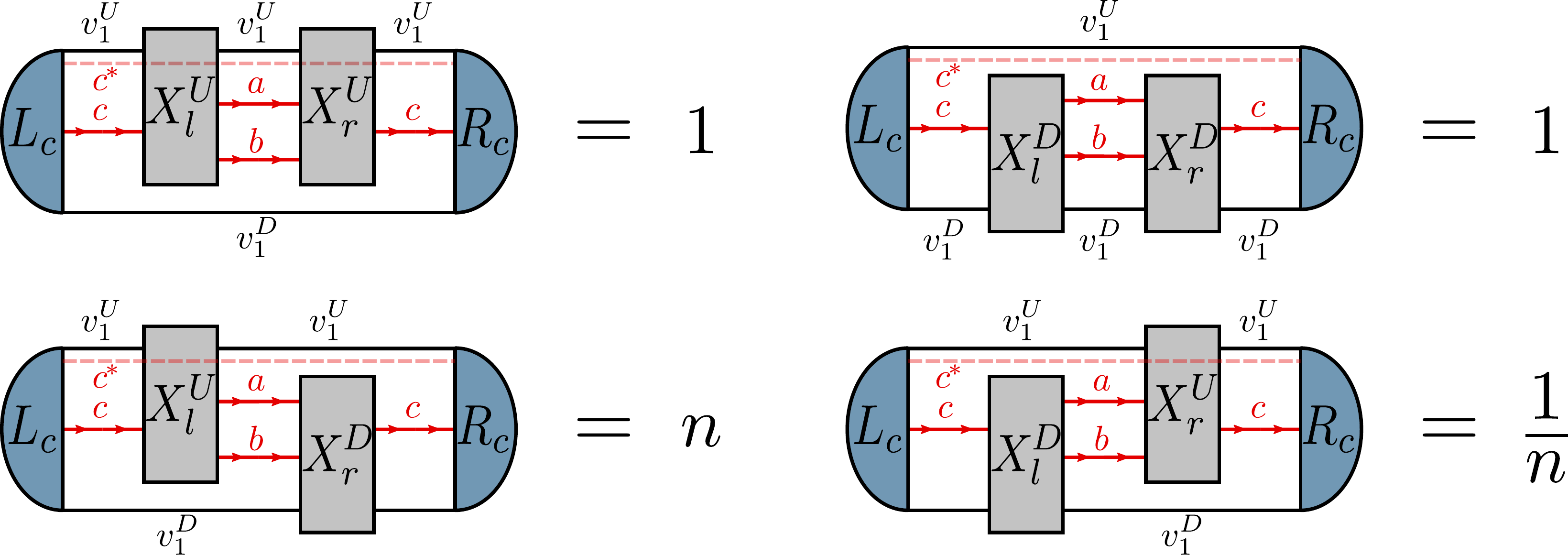}
\caption{ Normalization conditions for the $X_l$ and $X_r$ zippers. The projectors $X_l^U X_r^U$ and $X_l^D X_r^D$ in the top row act like identities when inserted between the left, $L_c$, and right, $R_c$, fixed points of the transfer matrix $Tr(v_1^U\otimes Z_c\otimes v_1^D\otimes Z_c^\dagger)$. Here the $Z_c^\dagger$, which is necessary for the diagram to be non-zero is represented by the dashed line. The mixed products $X_l^U X_r^D$ and $X_l^D X_r^U$ in the bottom row yield $n$ and $1/n$, respectively. The arbitrary $n$ can be brought to $1$ by rescaling the right $X$ zippers.}
\label{fig:Xconsistency}
\end{figure}

Having the full set of required $Y$-zippers $\lbrace Y_l^U,Y_r^U,Y_l^D,Y_r^D \rbrace$, we construct the $X$-zippers according to Fig. \ref{fig:Xzip}. From this construction we get that:
\begin{equation}
\sum\limits_{a,b}~(X^U_l)^c_{ab}~(X^U_r)^c_{ab} = (Y^U_r)^c_{1c}~(Y^U_l)^c_{1c},
\end{equation}
which is not necessarily equal to identity matrix $\mathbb{I}_{\chi_1\cdot \chi_c}$. However, it is a projector that acts like an identity when inserted between the left, $L_c$, and right, $R_c$, fixed points of the transfer matrix $Tr(v_1^U\otimes Z_c\otimes v_1^D\otimes Z_c^\dagger)$, as shown in the top row of Fig. \ref{fig:Xconsistency} that includes also the complementary $X^D$ case. 

A similar normalization between $X^U_l$ and its corresponding $X^D_r$ is not automatic, see the bottom row of Fig. \ref{fig:Xconsistency}. Here the number $n$ depends on somewhat arbitrary normalization of $Y$ zippers making the $X$ zippers. The number can be brought to $1$ by rescaling, e.g., $X^D_l\to n X^D_l$ and $X^D_r\to (1/n) X^D_r$. 
Having thus properly normalized all of the $X$-zippers we can proceed with the calculation of the $F$-symbols.

\section{Numerical $F$ symbols}
\label{sec:F}

The last step of the algorithm is to calculate the $F$-symbols in the equation:
\begin{figure}[H]
\includegraphics[scale=0.4]{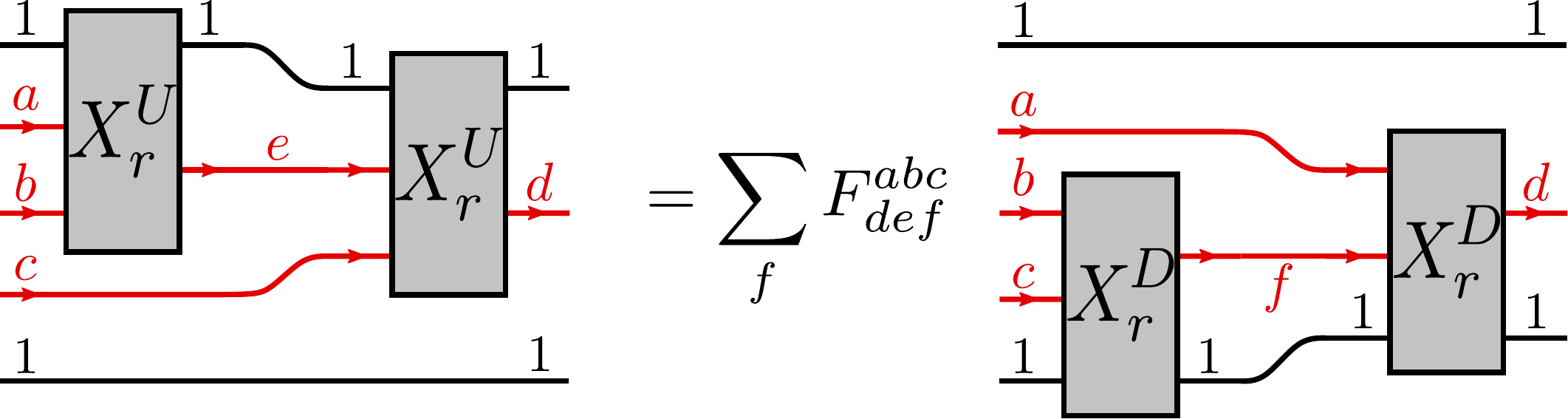}
\end{figure}
This is a coupled equation for $F^{abc}_{def}$ with different index $f$. In order to decouple it we project both sides onto $\big((X_l^D)^d_{ag}\cdot (X_l^D)^g_{bc}~ \big{\vert}$ from the left. At this point we verify that $\left( (X_l^D )^d_{ag}\cdot (X_l^D)^g_{bc}~ \big\vert ~ (X_r^D)^g_{bc} \cdot (X_r^D)^d_{ag} \right) = \delta_{gf}$ and we obtain an explicit formula:
\begin{figure}[H]
\includegraphics[scale=0.35]{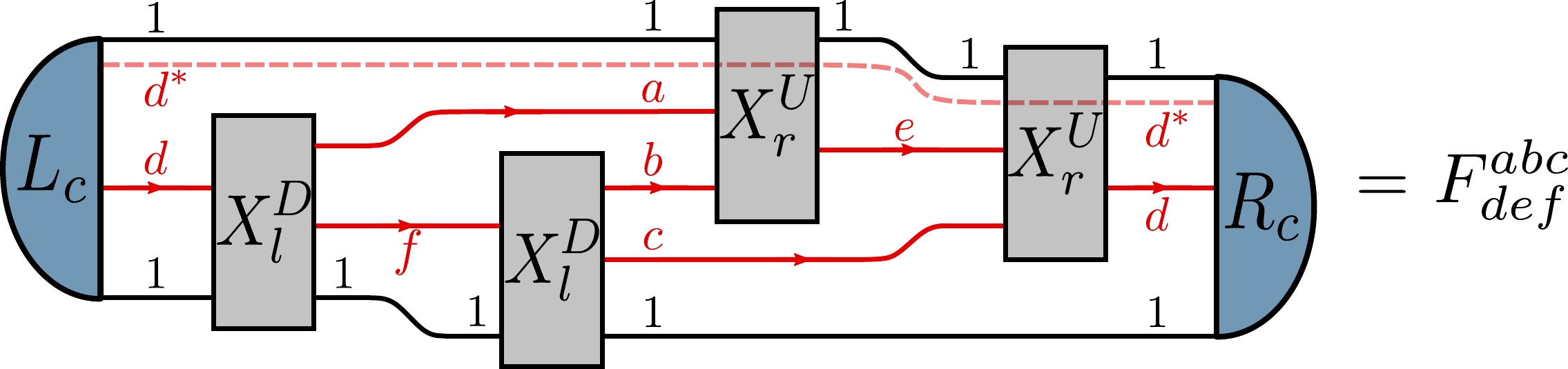}
\end{figure}
Here we have immersed the equation in the environment of left $L_c$ and right $R_c$ fixed points of the transfer matrix $Tr(v_1^U\otimes Z_d\otimes v_1^D\otimes Z_d^\dagger)$, the same as was used to find relative normalization of $X^U$ and $X^D$ zippers. The dotted red line denoted by $d^*$ is the trace over indices corresponding to $Z_d^\dagger$. 

A similar formula for an inverse of the matrix $F$ is
\begin{figure}[H]
\includegraphics[scale=0.35]{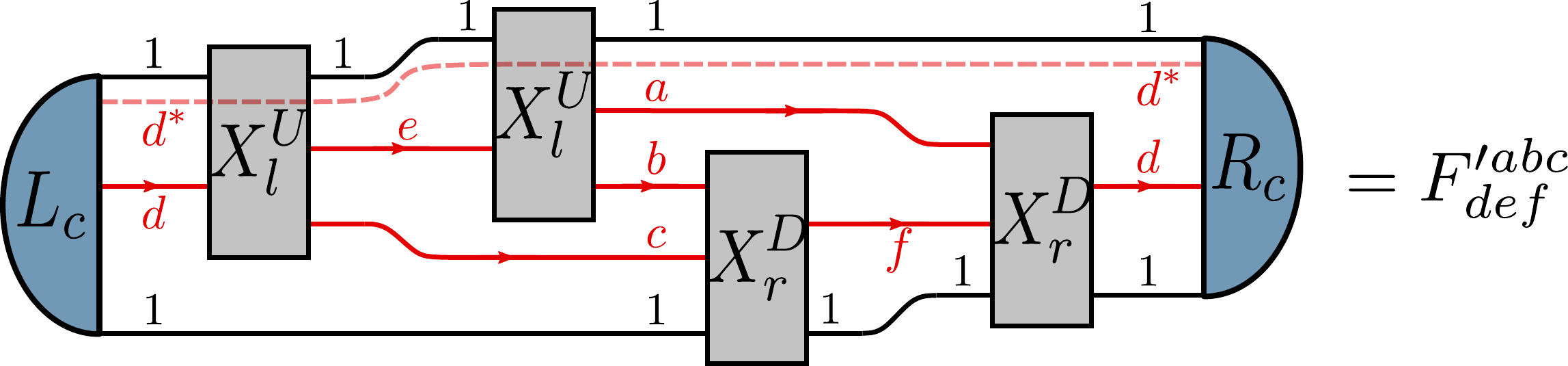}
\end{figure}
Both $F$ and $F'$ satisfy the Pentagon equation:
\begin{equation}
\sum_f F^{abc}_{def}\cdot  F^{b c i}_{h f j}\cdot F^{a f i}_{g d h} = F^{a b j}_{g e h}\cdot F^{e c i}_{g d j},
\label{eq:Pentagon}
\end{equation}
and describe the same topological order, although the value of their elements are in general different. The difference is manifestation of ``gauge freedom'' of  $F^{abc}_{def}$ due to remaining freedom in normalization of $X$ zippers:
\begin{eqnarray}
\lbrace X^U_l, X^U_r, X^D_l, X^D_r\rbrace ~\rightarrow~
\lbrace \lambda X^U_l, \frac{1}{\lambda} X^U_r, \lambda X^D_l, \frac{1}{\lambda} X^D_r\rbrace.\nonumber \\
\label{eq:gauge}
\end{eqnarray}
Here arbitrary $\lambda^c_{ab} \in\mathbb{C}$ depend on the labels of $X = X^c_{ab}$. Their values cannot be fixed by the Pentagon equation. $\lambda^c_{ab}$ parametrize gauge freedom of the $F$ symbols:
\begin{equation}
    F^{abc}_{def}  ~ \rightarrow ~ \frac{\lambda^f_{bc}\lambda^d_{af}}{\lambda^e_{ab}\lambda^d_{ec}} F^{abc}_{def},~~
    F'^{abc}_{def} ~ \rightarrow ~ \frac{\lambda^e_{ab}\lambda^d_{ec}}{\lambda^f_{bc}\lambda^d_{af}} F'^{abc}_{def}.
\label{eq:gaugeF}
\end{equation}
A straightforward way to proceed is to look for a gauge $\lambda^c_{ab}$ that brings the $F$ symbols, within numerical error, to a textbook form characteristic for a given type of topological order. This is what we do in the next section. 

It is important to point out that the topological order is given by the monoidal center $Z(\mC)$, meaning that two fusion categories $\mC_1$ and $\mC_2$ that a priori look completely different may describe the same topological order \cite{Lootens_2021}, in which case $\mC_1$ and $\mC_2$ are said to be \emph{Morita equivalent}. In order to deal with this redundancy, we compute the monoidal center in section \ref{sec:FtoST}, as well as the corresponding $S$ and $T$ matrices. 

\section{$F$ symbols in canonical gauge}
\label{sec:Fcanonical}

We use the gauge freedom in \eqref{eq:gaugeF} to bring $F$ symbols to a canonical gauge where, for a unitary fusion category, the matrices $F^{abc}_d$ are unitary and most elements of $F$ are one, especially if any of $a,b,c,d$ is trivial. To begin we notice that Eq. \eqref{eq:gaugeF} implies that a product
\begin{equation}
  F^{abc}_{def} F'^{abc}_{def},
  \label{FF'}
\end{equation} 
is gauge invariant. Therefore its square root will be used later to eliminate some of the gauge freedom of the $F$-symbols.

Additionally, all $X^c_{ab}$ where either $a=1$ ($b=1$) are chosen as identities between $b$ and $c$ ($a$ and $c$). This choice fixes the gauge partially as $\lambda^c_{1c} = 1 = \lambda^c_{c1}$ but we are still left with freedom to choose the $\lambda^c_{ab}$ where both $a\neq1$ and $b\neq1$. This residual gauge freedom leaves invariant all the $F^{abc}_{def}$ where one of $a,b,c$ is equal to $1$. 

Moreover, when there are only 2 anyon types in the input category, then also the $F^{abc}_{1ef}$ are left invariant by the residual gauge transformation. The only $F$-symbols that transform in a non-trivial way are:
\begin{eqnarray}
F^{222}_{212} &\rightarrow & 
\frac{\lambda^1_{22}}{\lambda^2_{22}\lambda^2_{22}}F^{222}_{212} \equiv \mu F^{222}_{212}, \\
F^{222}_{221} &\rightarrow & 
\frac{\lambda^2_{22}\lambda^2_{22}}{\lambda^1_{22}}F^{222}_{221} \equiv \frac{1}{\mu} F^{222}_{221}.
\label{eq:gaugen2}
\end{eqnarray}
In the unitary gauge for every fixed set of indices $a,b,c,d$ the matrix $F^{abc}_{def}$ is unitary in indices $ef$. We can choose $|\mu|$ such that magnitudes of $F^{222}_{2ef}$ become the same as square roots of corresponding products in \eqref{FF'}. With a proper phase of $\mu$ the matrix $F^{222}_{2ef}$ can be made unitary making manifest that the obtained $F$ symbols describe unitary fusion category.

When there are more than $2$ anyon types then there is a freedom:
\begin{equation}
    F^{abc}_{1ef} = F^{abc}_{1ca} \rightarrow \frac{\lambda^c_{ab} \lambda^1_{cc}}{\lambda^a_{bc}\lambda^1_{aa}} F^{abc}_{1ca}
\end{equation}
For $a=c$ this freedom is given by a simple ratio:
\begin{equation}
    F^{aba}_{1aa} \rightarrow \frac{\lambda^a_{ab} }{\lambda^a_{ba}} F^{aba}_{1aa} \equiv \mu(a,b) F^{aba}_{1aa}
\end{equation}
which allows to determine first non-trivial gauge transformation and eliminate it from all $F^{abc}_{def}$ in which it appears, by a substitution $\lambda^a_{ab}\rightarrow\mu(a,b)\lambda^a_{ba}$. The remaining scheme is largely model-dependent, but the general idea is to replace unknown $\lambda$'s with known ratios $\mu$ as we present on the examples below.

\subsection{Toric code and double semions} 

We obtain the $Y$ and $X$ zipper tensors. For those 2 Abelian models there are 2 trivial $Y$-zippers: $Y^1_{11}, Y^2_{21}=\mathbb{I}$ and 2 non-trivial $Y$-zippers: $Y^2_{12}, Y^1_{22}$ giving rise to 4 non-zero $X$-zippers: $X^1_{11}, X^2_{12}, X^2_{21}, X^1_{22}$. For both the toric code and the double semion model all gauges $\lambda$ in Eq. \eqref{eq:gauge} cancel each other in the expressions for $F$ symbols. We obtain numerically exact $F$-symbols immediately in the canonical gauge for toric code:
\begin{equation}
    F^{abc}_{def} = N^e_{ab}N^e_{cd}N^f_{ad}N^f_{bc}
\end{equation} 
and the same for double semions with the exception for $F^{222}_{211}=-1$.

\subsection{Twisted quantum double of $Z_3$}
We find 3 trivial Y zippers with both up and down eigenvectors $Y^i_{i1}$ for $i=1,q,q^*$ and 6 nontrivial with up-eigenvectors: 
\begin{equation}
    (Y^U)^2_{12},(Y^U)^3_{13},(Y^U)^1_{23},(Y^U)^1_{32},(Y^U)^3_{22},(Y^U)^2_{33}
\end{equation} 
and with down eigenvectors:
\begin{equation}
(Y^D)^3_{12},(Y^D)^2_{13},(Y^D)^3_{23},(Y^D)^2_{32},(Y^D)^1_{22},(Y^D)^1_{33},
\end{equation}
which altogether give rise to a unique set of $X$ zippers: trivial $X^a_{1a}, X^a_{a1}$ for $a=1,q,q^*$ and $X^3_{22}, X^2_{33}, X^1_{23}, X^1_{32}$. 
Therefore there are 4 random residual gauges: $\lambda^1_{23}, \lambda^1_{32}, \lambda^3_{22}, \lambda^2_{33}$, which appear in only 2 combinations:
\begin{equation}
    \rho_1 = \frac{\lambda^1_{23}}{\lambda^1_{32}}, ~\rho_2 = \frac{\lambda^1_{32}}{\lambda^3_{22}\lambda^2_{33}},
\end{equation}
where $\rho_1$ and its inverse fully fixes $F^{222}_{133}, F^{323}_{311}, F^{232}_{211}, F^{333}_{122}$ while $\rho_2$ fully fixes $F^{332}_{321},F^{322}_{213}$ and there are only two remaining $F$-symbols:
\begin{eqnarray}
F^{223}_{231} \rightarrow \rho_1\rho_2 ~F^{223}_{231} \nonumber \\
F^{233}_{312} \rightarrow \frac{1}{\rho_1\rho_2}~ F^{233}_{312}.
\end{eqnarray}
This procedure allows us to obtain $F^{abc}_{def} = N^e_{ab}N^e_{cd}N^f_{ad}N^f_{bc}$ with the exception of 
\begin{equation}
    F^{222}_{133} = F^{333}_{122} = (F^{332}_{321})^* = (F^{223}_{231})^* = \mathrm{e}^{\frac{2i\pi}{3}}.
\end{equation}
The obtained $F$-symbols necessarily satisfy the Pentagon equation, both before and after the gauge transformation. This is the only example we present, in which the $F$-symbols and their inverses $F^{-1}$ are not equal, but actually $(F^{abc}_{def})^{-1} = (F^{abc}_{def})^*$.

\subsection{Fibonacci string net} 

We obtain 2 trivial $Y$-zippers: $Y^1_{11}, Y^2_{21}=\mathbb{I}$ and 3 non-trivial $Y$-zippers: $Y^2_{12}, Y^1_{22}, Y^2_{22}$ giving rise to 5 non-zero $X$-zippers: $X^1_{11}, X^2_{12}, X^2_{21}, X^1_{22}, X^2_{22}$. With $X$-zippers we obtain $F$ symbols that satisfy the Pentagon equation within machine precision. However, the obtained $F$ symbols turn out to be in a random non-unitary gauge.

As the double Fibonacci model has 2 anyon types, the residual gauge freedom in Eq. \ref{eq:gaugen2} can be employed to adjust both $F^{abc}_{def}$ or $F'^{abc}_{def}$ to the absolute values obtained from a square root of the product \eqref{FF'} and then to fix their phase in such a way that $F^{abc}_{def}$ and $F'^{abc}_{def}$ become unitary in indices $e,f$ within numerical precision. This way we obtain $F^{abc}_{def} = N^e_{ab}N^e_{cd}N^f_{ad}N^f_{bc}$ except for
\begin{eqnarray}
    F^{\tau\tau\tau}_{\tau11} = -F^{\tau\tau\tau}_{\tau\tau\tau} = \frac{1}{d_\tau}, ~~
    F^{\tau\tau\tau}_{\tau\tau1} = F^{\tau\tau\tau}_{\tau1\tau} = \frac{1}{\sqrt{d_\tau}}.
\end{eqnarray}
Here the quantum dimension $d_\tau=(\sqrt5+1)/2$.

\subsection{Fibonacci string net with local filtering} 

For the local filtering that introduces a finite correlation length $\xi$ and drives the state away from the fixed point we used the same algorithm. Accuracy of the obtained $F$ symbols, measured by the Frobenius norm $\epsilon_F = ||F_{numerical}-F_{canonical}||$, with $F_{canonical}$ being the exact solution, is listed in the table \ref{tab:betaxi2}. The accuracy is still high for small perturbations and we leave the improvement of this algorithm to achieve better results for higher correlation lengths for future work.

\begin{table}[H]
\centering \begin{tabular}{c|c|c}
$\beta$ & $\xi$ &      $\epsilon_F$        \\ \hline
0.01    & 0.23  &  $\mathcal{O}(10^{-7})$  \\
0.05    & 0.42  &  $\mathcal{O}(10^{-5})$  \\
0.12    & 1.04  &  $ \mathcal{O}(10^{-2})$ \\ 
0.15    & 2.32  &  $\mathcal{O}(10^{-2})$  
\end{tabular}
\caption{The error of the obtained $F$ symbols, calculated as the Frobenius norm of the difference with respect to the exact ones, calculated in a paragraph above. First column represents the perturbation strength $\beta$, second -- the correlation length $\xi$ and the errors of the $F^{abc}_{def}$ are shown in the last column.}
\label{tab:betaxi2}
\end{table}

\subsection{Ising string net}
As the double Ising model has 3 anyon types in the input category $\mC$ there are 5 random residual gauges: $\lambda^1_{\psi\psi},\lambda^1_{\sigma\sigma},\lambda^\sigma_{\psi\sigma},\lambda^\sigma_{\sigma\psi},\lambda^\psi_{\sigma\sigma}$. They appear in only 3 combinations:
\begin{equation}
\rho_1 = \frac{\lambda^\sigma_{\psi\sigma}}{\lambda^\sigma_{\sigma\psi}},~~
\rho_2 = \frac{\lambda^1_{\sigma\sigma}}{\lambda^\sigma_{\sigma\psi}\lambda^\psi_{\sigma\sigma}},~~
\rho_3 = \frac{\lambda^1_{\psi\psi}}{(\lambda^\sigma_{\sigma\psi})^2}.
\end{equation}
Starting with $\rho_1$, which fully fixes $F^{\sigma\psi\sigma}_{1\sigma\sigma}, F^{\sigma\psi\sigma}_{\psi\sigma\sigma}$ and $F^{\sigma\sigma\sigma}_{\sigma\psi\psi}$, we find $\rho_2$ fixing $F^{\sigma\sigma\psi}_{\psi1\sigma}, F^{\sigma\sigma\sigma}_{\sigma1\psi}$, and finally $\rho_3$ which is fully fixing $F^{\sigma\psi\psi}_{\sigma\sigma1}$. The remaining $F$ symbols are fixed by proper combinations:
\begin{eqnarray}
F^{\psi\psi\sigma}_{\sigma1\sigma} &\rightarrow& \frac{\rho_3}{\rho_1^2}~F^{\psi\psi\sigma}_{\sigma1\sigma} \nonumber \\
F^{\psi\sigma\sigma}_{1\sigma\psi} &\rightarrow& \frac{\rho_1\cdot\rho_2}{\rho_3}~F^{\psi\sigma\sigma}_{1\sigma\psi} \nonumber \\
F^{\psi\sigma\sigma}_{\psi\sigma1} &\rightarrow& \frac{\rho_1}{\rho_2}~F^{\psi\sigma\sigma}_{\psi\sigma1} \nonumber \\
F^{\sigma\sigma\sigma}_{\sigma\psi1} &\rightarrow& \frac{\rho_1}{\rho_2}~F^{\sigma\sigma\sigma}_{\sigma\psi1} \nonumber \\
F^{\sigma\sigma\psi}_{1\psi\sigma} &\rightarrow& \frac{\rho_3}{\rho_2}~F^{\sigma\sigma\psi}_{1\psi\sigma} 
\end{eqnarray}
This way we obtain $F^{abc}_{def} = N^e_{ab}N^e_{cd}N^f_{ad}N^f_{bc}$ except for
\begin{eqnarray}
  &&
  F^{\sigma\sigma\sigma}_{\sigma11} = F^{\sigma\sigma\sigma}_{\sigma1\psi} = F^{\sigma\sigma\sigma}_{\sigma\psi1} = -F^{\sigma\sigma\sigma}_{\sigma\psi\psi} = \frac{1}{\sqrt{2}}, \\
  &&
  F^{\psi\sigma\psi}_{\sigma\sigma\sigma} = F^{\sigma\psi\sigma}_{\psi\sigma\sigma} = -1,
\end{eqnarray}
all with numerical precision.

\subsection{Quantum double of $S_3$}
\subsubsection*{$\text{Vec}(S_3)$ MPO symmetries}
Apart from the trivial $X$-zippers with an identity symmetry $X^a_{1a} = X^a_{a1} = \mathbb{I}_a$ there are 6 nontrivial ones $X^1_{22},X^1_{33},X^2_{22},X^3_{22},X^2_{32},X^2_{23}$, all with its corresponding gauge-freedom $\lambda^c_{ab}$. However there are only 4 independent variables:
\begin{equation}
    \rho_1 = \frac{\lambda^2_{23}}{\lambda^2_{32}}, \rho_2 = \frac{(\lambda^2_{22})^2}{\lambda^1_{22}}, \rho_3 = \frac{\lambda^1_{22}}{\lambda^3_{22}\lambda^1_{33}}, \rho_4 = \frac{\lambda^1_{33}}{(\lambda^3_{22})^2}.
\end{equation}
After elimination of the gauge freedom  from all possible $F$-symbols containing the aforementioned ratios we obtain that $F^{abc}_{def} = N^e_{ab}N^e_{cd}N^f_{ad}N^f_{bc}$, except for:
\begin{eqnarray}
    &F^{322}_{222} = F^{232}_{222} = F^{223}_{222} = F^{222}_{322} = -1 &\nonumber \\
    &F^{222}_{211} = F^{222}_{231} = F^{222}_{213} = F^{222}_{223} = \frac{1}{d_\pi} &\nonumber \\
    &F^{222}_{221} = F^{222}_{212} = -F^{222}_{232} = -F^{222}_{223} = \frac{1}{\sqrt{d_\pi}} &\nonumber \\
    &F^{222}_{222} = 0,&
\end{eqnarray}
where $d_\pi = 2$ is the quantum dimension of $\pi$ and the remaining quantum dimensions are $d_1=d_\psi=1$.

\subsubsection*{$\text{Rep}(S_3)$ MPO symmetries}
There are 25 nontrivial fusion rules $N^c_{ab}$ giving rise to corresponding $X$ zippers $X^c_{ab}$, hence 25 random gauges $\lambda^c_{ab}$, which can be eliminated using only 20 ratios $\rho_i$. It can be done by subsequent substitution of certain ratios $\rho_i = \frac{\lambda^e_{ab}\lambda^d_{ec}}{\lambda^f_{bc}\lambda^d_{af}}$, so that the final $F$-symbols are all trivial: $F^{abc}_{def} = N^e_{ab}N^e_{cd}N^f_{ad}N^f_{bc}$, with every index taking up to 6 values. All the quantum dimensions are $d_a = 1$.

At first glance the $F$-symbols in both examples above may seem to describe completely different topological orders as they describe different unitary fusion categories UFC. However the calculation of the Drinfeld center in the following sections proves that this is not the case.


\section{ $S$ and $T$ matrices from $F$ symbols}
\label{sec:FtoST}
The topological $S$ and $T$ matrices are gauge invariant quantities, which in principle could be obtained from the $F$ symbols in arbitrary gauge by considering proper gauge-cancelling factors\cite{WenQalgebra}. Here instead, we make use of the $F$-symbols in canonical gauge, obtained in section \ref{sec:Fcanonical}, to derive a simpler expression.

An important observation is that the labels of all non-zero elements of both $X$ and $Y$ zippers define the possible fusions $N^c_{ab}$ of the anyons in the category, from which we obtain their quantum dimensions $d_a$, as the largest magnitude eigenvalue of the $N_a$ matrix. In this sense fusion rules and quantum dimensions are exact independently of the correlations in the models. 

In order to obtain all the anyons or definite anyonic sectors (MES) in the tensor network ansatz we need to find central idempotents of the algebra generated by elements $A_{abcd}\propto  N^b_{da}N^b_{cd} \propto (X_r)^b_{da} (X_l)^b_{cd}$ (connected through the index b, but not summed over b), where we omit possible multiplicities as they are all equal 1 in our examples. Central idempotents, when inserted into PEPS, can be thought of as projectors onto states with well-defined anyon flux along the torus. The multiplication of the basis elements $e_i:=A_{abcd}$ defines some algebra, from which we find both central and simple idempotents as desrcibed in Appendix \ref{app:CI}.
The algebra of $A_{abcd}$ can be used to calculate the action of the Dehn twist on a state with a symmetry $Z_a$ along the torus\cite{williamson2017SET}:
\begin{eqnarray}
\tilde{T}(A_{abad}) &=& A_{a1a\bar{a}}\cdot A_{abad} \nonumber \\
 &=&\sum_{e,c} \sqrt{\frac{d_a d_{\bar{a}} d_c d_d}{d_e d_b}} (F^{\bar{a}ad}_{d1b})^{-1} F^{\bar{a}da}_{deb} F^{a\bar{a}d}_{d1e} A_{acae}\delta_{cd}, \nonumber \\
e_i &=& \sum_{j} \tilde{T}_{ij}e_j
\label{eq:Tmatrix}
\end{eqnarray}
This formula gives rise to the $\tilde{T}$ matrix in the basis of $e_i \equiv A_{abcd}$. In the eigenbasis (the MES basis) this matrix is diagonal and contains the phases corresponding to topological spins: $T = \text{diag}(\theta_1,...\theta_{N})$. However, at this point we do not posses enough knowledge to assign anyon labels to them and certain topological spins belonging to multidimensional particles in non-Abelian anyon models are repeated (e.g. $\theta_{\tau\bar{\tau}}$ in the double Fibonacci string net and $\theta_{\sigma\bar{\sigma}}$ in the double Ising string net). Therefore we proceed with the calculation of the topological $S$ and $T$ matrices in the MES basis.  If we denote central idempotents inserted in PEPS to create a minimally entangled state in $y$-direction by $\mathcal{P}_i^y$ and similarly in the $x$-direction by $\mathcal{P}_i^x$, then the transformation between these two basis is actually an $S$ matrix:
\begin{equation}
    \mathcal{P}_i^y = \sum\limits_j S_{ij} \mathcal{P}_j^x
\end{equation}
We can further write this expression in terms of the basis elements $e_k := A_{abcd}$:
\begin{equation}
    \mathcal{P}_i^y = \sum_a c^i_a e^y_a = \sum_j S_{ij} \sum_b c^j_b e^x_b,
\end{equation}
which written in the matrix forms without summations, with $E^{x,y}$ being the basis in $x$ and $y$ respectively, $B$ the basis change between $x$ and $y$, $P$ - the matrix of coefficients of the central idempotents in the $E$ basis, is:
\begin{equation}
    P E^y = P B E^x = S P E^x, ~\Rightarrow~ S = P B P^{-1}
\end{equation}
Similarly we obtain the expression for the $T$ matrix in the MES basis:
\begin{equation}
T = P \tilde{T} P^{-1}
\end{equation}
The $\tilde{T}$ was is given in eq.\ref{eq:Tmatrix} and the basis change $B$ is given by the combination of F-symbols, as shown in\cite{williamson2017SET}:
\begin{equation}
    S(A_{abad}) = \sum_e  d_a \sqrt{\frac{d_d d_{\bar{d}}}{d_ed_b} } (F^{ad\bar{d}}_{ab1})^* F^{d\bar{d}a}_{a1e} F^{da\bar{d}}_{abe} A_{\bar{d}e\bar{d}a}~
\end{equation}
For non-Abelian anyon models, the inversion $P^{-1}$ for 2 or more dimensional idempotents actually means the sum of inverted simple idempotents. Unlike the matrix of central idempotent, the matrix of simple idempotents $P_{\text{simple}}$ in most cases is square and invertible. Technically it means that the matrix $P_{\text{simple}}$ is made of rows of all simple idempotents, which makes it block-diagonal with two (or more) rows in different blocks corresponding to the same anyon type.  Next we invert the matrix of simple idempotents $P_{\text{simple}}$, so that the columns of $P_{\text{simple}}^{-1}$ correspond to the inverses of simple idempotents. In the end we sum up the columns that correspond to the same anyon flux to get $P^{-1}$. Moreover the rows of $P$ corresponding to anyon types that are supported on this multidimensional spaces have to be normalized (divided by their dimensionality).

\subsection{Toric code, double semion, Fibonacci and Ising string net}

For all the RG fixed point wave-functions of toric code, double semion, double Fibonacci and double Ising we obtain correct topological $S$ and $T$ matrices within machine precision. All the results are listed below.
\begin{itemize}
    \item Toric code
    \begin{equation*}
{\scriptsize S_{\rm{TC}} = \frac{1}{2}
\begin{pmatrix}
  1 & 1 & 1 & 1\\
  1 & 1 & -1 & -1\\
  1 & -1& 1& -1 \\
  1 & -1 & -1 & 1 \\
 \end{pmatrix},\qquad
 \scriptsize
  T_{\rm{TC}} = \begin{pmatrix}
  1 & 0 & 0 & 0 \\
  0 & 1 & 0 & 0 \\
  0 & 0 & 1 & 0 \\
  0 & 0 & 0 & -1 \\
 \end{pmatrix}} \ .
\end{equation*}

\item double semion
\begin{align*}
S_{\rm{ds}} = \frac{1}{2}
\begin{pmatrix}
1 & 1 \\
1 & -1\\
\end{pmatrix}^{\otimes 2},\qquad
T_{\rm{ds}} = \begin{pmatrix}
1 & 0 \\
0 & i \\
\end{pmatrix} \otimes
\begin{pmatrix}
1 & 0 \\
0 & -i \\
\end{pmatrix}.
\end{align*}

\item double Fibonacci, with $\varphi=\frac{1+\sqrt{2}}{2}$

\begin{align*}
S_{\rm dFib}&=\frac{1}{\varphi+2}
    \begin{pmatrix}
    1       &  \varphi\\
    \varphi  & -1\\
    \end{pmatrix}^{\otimes 2},\\[1em]
T_{\rm dFib} &=\begin{pmatrix}
1 & 0 \\
0 & \mathrm{e}^\frac{4i\pi}{5}\\
\end{pmatrix} \otimes 
\begin{pmatrix}
1 & 0 \\
0 & \mathrm{e}^\frac{-4i\pi}{5}\\
\end{pmatrix}
.
\end{align*}

\item double Ising

\begin{align*}
S_{\rm{dIs}} &= \frac{1}{4} \begin{pmatrix}
1 & \sqrt{2} & 1\\
\sqrt{2} & 0 & -\sqrt{2}\\
1 & -\sqrt{2} & 1
\end{pmatrix}^{\otimes 2},\\[1em]
T_{\rm{dIs}} &= \begin{pmatrix}
1 & 0 & 0\\
0 & \mathrm{e}^\frac{i\pi}{8} & 0\\
0 & 0 & -1
\end{pmatrix} \otimes \begin{pmatrix}
1 & 0 & 0\\
0 & \mathrm{e}^\frac{-i\pi}{8} & 0\\
0 & 0 & -1
\end{pmatrix}.
\end{align*}
\end{itemize}

\subsection{Fibonacci string-net with local filtering}

By direct application of the described procedure for perturbations $\beta=0.01,~0.05$ we can recover topological modular matrices with satisfying precision as shown in the table \ref{tab:ST}. For higher $\beta$ in order to obtain the $S_{\rm Fib}$ and $T_{\rm Fib}$ matrices we need to improve the quality of $F^{abc}_{def}$ to satisfy the Pentagon equation \ref{eq:Pentagon} with better accuracy. Here we perform a simple Monte Carlo (MC), where we sweep over all non-zero elements of $F$ tensor, apart from $F^{111}_{111}$, $F^{211}_{221}$, $F^{121}_{222}$, $F^{112}_{212}$, which are all equal 1 by construction. In a single MC move we change an element of $F$ tensor $F'^{abc}_{def} = F^{abc}_{def} + \delta\cdot r_1$, where $r_1$ is a random complex number and $\delta = 0.01\cdot\epsilon(F)$ is the MC step with $\epsilon(F)$ being the error of the Pentagon equation. We calculate the new error of the Pentagon equation $\epsilon(F')$ and accept it if $\epsilon(F')<\epsilon(F)$ or check if the ratio $\frac{\epsilon(F)}{\epsilon(F')}$ is smaller than another random real number $r_2$ and accept the move if this is fulfilled. We perform such sweeps over all aforementioned elements of $F$ tensor, which enables to obtain an error low enough to calculate the topological $S$ and $T$ matrices. 
We list the new error of $F$ symbols together with the errors for the topological $S$ and $T$ matrices in table \ref{tab:ST}.

\subsection{Twisted quantum double of $\mathbb{Z}_3$}
For the twisted quantum double of $\mathbb{Z}_3$ we obtain the following $T$ and $S$ matrices:
\begin{align*}
\text{diag}(T_{\mathbb{Z}_3}) &= { \left(1,1,1,\mathrm{e}^{\frac{4i\pi}{9}},\mathrm{e}^{\frac{-8i\pi}{9}},\mathrm{e}^{\frac{-2i\pi}{9}},\mathrm{e}^{\frac{-2i\pi}{9}},\mathrm{e}^{\frac{-8i\pi}{9}},\mathrm{e}^{\frac{4i\pi}{9}}\right), } \\[1em]
\frac{\arg(S_{\mathbb{Z}_3})}{2\pi} &= {\begin{pmatrix}
0 & 0 & 0 & 0 & 0 & 0 & 0 & 0 & 0 \\
0 & 0 & 0 & -\frac{1}{3} & -\frac{1}{3} & -\frac{1}{3} & \frac{1}{3} & \frac{1}{3} & \frac{1}{3}\\
0 & 0 & 0 & \frac{1}{3} & \frac{1}{3} & \frac{1}{3}  & -\frac{1}{3} & -\frac{1}{3} & -\frac{1}{3}\\
0 & -\frac{1}{3} & \frac{1}{3} & -\frac{4}{9} & \frac{2}{9} & -\frac{1}{9} & \frac{1}{9} & -\frac{2}{9} & \frac{4}{9}\\
0 & -\frac{1}{3} & \frac{1}{3} & \frac{2}{9} & -\frac{1}{9} & -\frac{4}{9} & \frac{4}{9} & \frac{1}{9} & -\frac{2}{9}\\
0 & -\frac{1}{3} & \frac{1}{3} & -\frac{1}{9} & -\frac{4}{9} & \frac{2}{9} & -\frac{2}{9} & \frac{4}{9} & \frac{1}{9}\\
0 & \frac{1}{3}  & -\frac{1}{3} & \frac{1}{9} & \frac{4}{9} & -\frac{2}{9} & \frac{2}{9} & -\frac{4}{9} & -\frac{1}{9}\\
0 & \frac{1}{3}  & -\frac{1}{3} & -\frac{2}{9} & \frac{1}{9} & \frac{4}{9} & -\frac{4}{9} & -\frac{1}{9} & \frac{2}{9} \\
0 & \frac{1}{3}  & -\frac{1}{3} & \frac{4}{9} & -\frac{2}{9} & \frac{1}{9} & -\frac{1}{9} & \frac{2}{9} & -\frac{4}{9}
\end{pmatrix},}\\[1em]
|(S_{\mathbb{Z}_3})_{ij}| &= \frac{1}{3}.
\end{align*}


\begin{table}[t]
\centering \begin{tabular}{c|c|c|c|c|c}
$\beta$ & $\xi$ &  $\chi$ &  $\epsilon_S$     & $\epsilon_T$ & $\epsilon_F^{MC}$     \\ \hline
0.01    & 0.23  &  8      &  $\mathcal{O}(10^{-7})$ & $\mathcal{O}(10^{-7})$ & --- \\
0.05    & 0.42  &  12     &  $\mathcal{O}(10^{-5})$ & $\mathcal{O}(10^{-5})$ & --- \\ \hline
0.12    & 1.04  &  16     &  $\mathcal{O}(10^{-3})$ & $\mathcal{O}(10^{-6})$ & $\mathcal{O}(10^{-3})$ \\
0.15    & 2.32  &  20      &  $\mathcal{O}(10^{-3})$ & $\mathcal{O}(10^{-6})$ & $\mathcal{O}(10^{-3})$
\end{tabular}
\caption{The error of the final $S$ and $T$ matrices, $\epsilon_S$ and $\epsilon_T$ respectively, calculated as the Frobenius norm of the difference between the numerical and the exact ones for different perturbation strengths $\beta$. Here $\chi$ is the bond dimension of the boundary eigenvectors $v_i^{U,D}$ used for the calculations. The numerical $S$ and $T$ matrices for bigger perturbations $\beta = 0.12, 0.15$ can be obtained only after the error of the numerical $F$-symbols is reduced up to the value of $\epsilon^{MC}_F$ by simple Monte Carlo minimization of the error of Pentagon equation. }
\label{tab:ST}
\end{table}


\subsection{Quantum double of $S_3$}
For both iPEPS representations we obtain the topological $S$ and $T$ matrices which agree with machine precision with the exact ones up to the simultaneous permutation of columns and rows.
\begin{equation}
\begin{split}
\text{diag}(T_{S_3}) = {\scriptsize \left(1,\mathrm{e}^{\frac{-2i\cdot\pi}{3}},1,-1,\mathrm{e}^{\frac{2i\cdot\pi}{3}},1,1,1\right), }\\
S_{S_3} = {\scriptsize \frac{1}{6}\begin{pmatrix}
1 & 3 & 2 & 1 & 2 & 3 & 2 & 2 \\
3 & 3 & 0 & -3 & 0 & -3 & 0 & 0\\
2 & 0 & 4 & 2 & -2 & 0 & -2 & -2\\
1 & -3 & 2 & 1 & 2 & -3 & 2 & 2\\
2 & 0 & -2 & 2 & -2 & 0 & -2 & 4\\
3 & -3 & 0 & -3 & 0 & 3 & 0 & 0\\
2 & 0 & -2 & 2 & -2 & 0 & 4 & -2\\
2 & 0 & -2 & 2 & 4 & 0 & -2 & -2
\end{pmatrix}.}
\end{split}
\nonumber
\end{equation}
The algorithm we present in Appendix \ref{app:CI} fails to decompose the two-dimensional central idempotent corresponding to the anyon flux $(1,\pi)$ into simple idempotents, which should be done as shown in \cite{williamson2017SET}:
\begin{eqnarray}
    (1,\pi) &=& (1,\pi)_{00} + (1,\pi)_{11} \nonumber \\
    (1,\pi) &=& \frac{1}{3}\left( 2\cdot A_{1111} -A_{1515} - A_{1616} \right) \\
    (1,\pi)_{00} &=& \frac{1}{3}\left( A_{1111} + \mathrm{e}^{\frac{-2i\cdot\pi}{3}} A_{1515} + \mathrm{e}^{\frac{2i\cdot\pi}{3}} A_{1616}\right) \\
    (1,\pi)_{11} &=& \frac{1}{3}\left( A_{1111} + \mathrm{e}^{\frac{2i\cdot\pi}{3}} A_{1515} + \mathrm{e}^{\frac{-2i\cdot\pi}{3}} A_{1616}\right) 
\end{eqnarray}
However that is a necessary step to do in order to obtain correct modular $S$ and $T$ matrices shown above.

\section{Conclusion and outlook}
\label{sec:summary}

The numerical method to obtain the $F$-symbols of the fusion category fully characterizing the topological order can be summarized in the following few steps:
\begin{enumerate}
    \item Finding all boundary fixed points, both up and down, $v_i^U$, $v_i^D$ of the double iPEPS horizontal transfer matrix $\Omega_h$.
    \item Calculating all iMPO symmetries $Z_a$ mapping between different boundary fixed points: $v_i^U\cdot Z_a = v_j^u$ and $v_j^D\cdot Z_a^T = v_j^D$.
    \item Finding the gauge transformations $Y^k_{ia}$ between equal iMPOs $M = v_i \cdot Z_a$ and $v_k$ for both up and down eigenvectors and combining them to yield zippers $X^c_{ab}$ fusing the product of iMPO symmetries $v_1Z_a\cdot Z_b$ into single iMPO symmetry $v_1Z_c$.
    \item Calculating the $F$-symbols using the associativity of the fusions of iMPO symmetries $Z_a$.
    \item The numerical $F$-symbols in the random gauge can be brought into canonical gauge by proper inspection of the freedom in the normalization of $X^c_{ab}$ zipppers. They can also be used to calculate gauge invariant topological data in the form of $S$ and $T$ matrices encoding mutual and self statistics of the emergent anyons of the doubled category.
\end{enumerate}
After slight purification of the $F$-symbols, by minimizing the error of the pentagon equation, the method proved to give accurate results for states with correlation length up to $\xi=2.3$.

As mentioned in the introduction, by studying the virtual iMPO symmetries rather than the physical Wilson line operators, we avoid the complications that arise due to the broadening of these Wilson lines away from the fixed point. Although we have shown that the method is applicable with nonzero correlation length, the correlation lengths for which the correct results are recovered are still rather small and it is clear there is still much room for improvement in several aspects of the algorithm. 
	
Such improvements would allow us to study the change in topological order when driving a certain state through a phase transition. These phase transitions are characterized by the breaking and emergence of MPO symmetries, which should be reflected in the fixed point structure. This becomes particularly interesting when considering variationally optimized iPEPS, where a specific choice of PEPS representation and corresponding MPO symmetries is not imposed but rather chosen by the algorithm. Close to a phase transition, we expect the algorithm to prefer the PEPS representation that most naturally allows the relevant MPO symmetries to be broken or emerge.


\acknowledgements 

Numerical calculations were performed in MATLAB with the help of \verb+ncon+ function \cite{NCON} for tensor contractions. 
AF acknowledges financial support by Polish Ministry of Science and Education, project No. DI2015 021345, from the budget funds for science in 2016-2020 under the Diamond Grant program.
This research was also supported by Narodowe Centrum Nauki (NCN) under grant 2019/35/B/ST3/01028 
(AF, JD) and Etiuda grant 2020/36/T/ST3/00451 (AF). 
 This work has received funding from the European Research Council (ERC) under  the  European  Union’s  Horizon  2020  research  and  innovation  programme  (grant agreement No 715861 (ERQUAF) and 647905 (QUTE)). LL is supported by a PhD fellowship from the Research Foundation Flanders (FWO).


\bibliography{refs.bib}


\appendix



\begin{figure}[b]
\centering
\includegraphics[width=0.85\columnwidth]{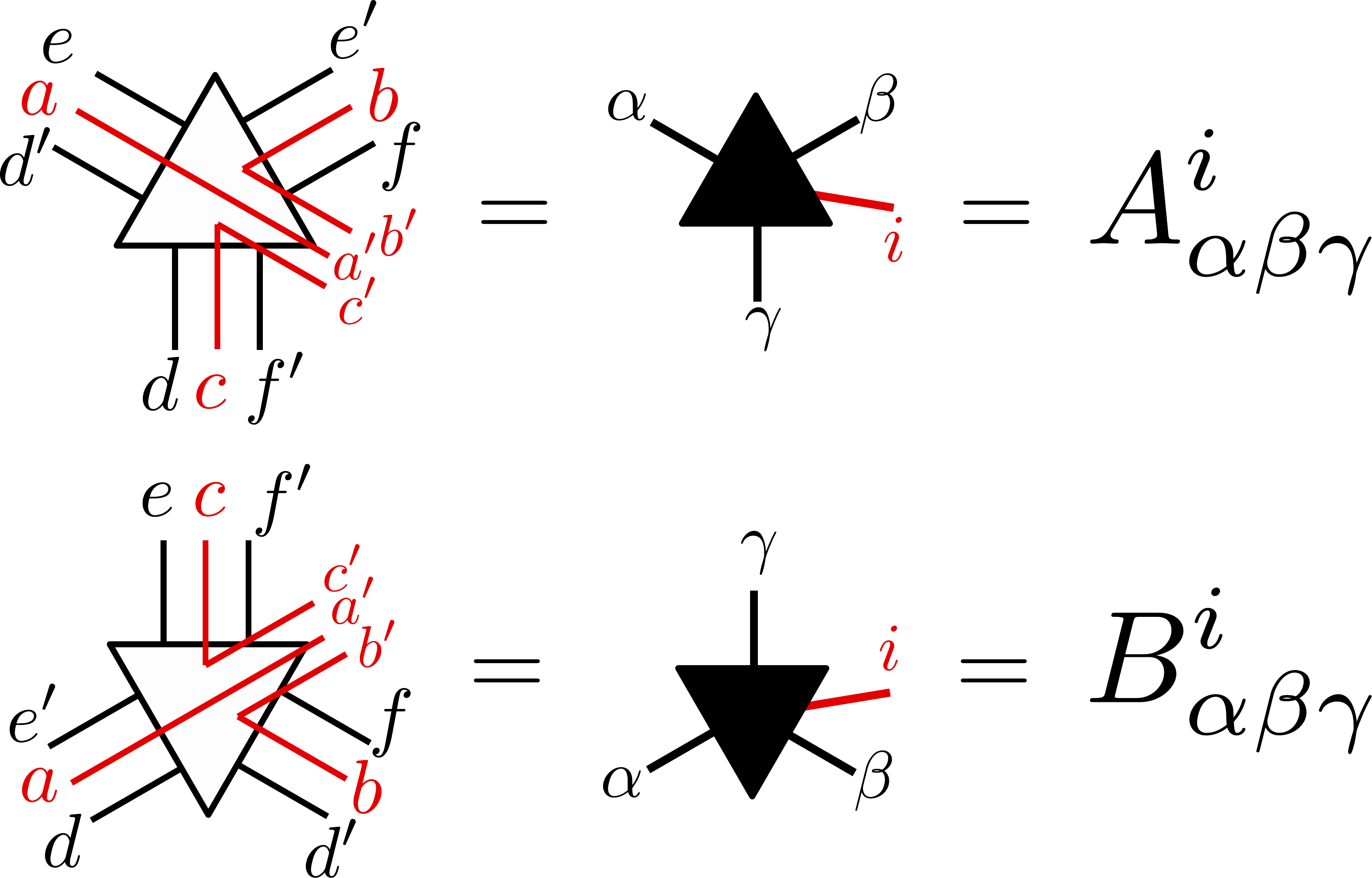}
\caption{
Tensors forming the iPEPS are defined via combination of $F$-symbols and corresponding quantum dimensions $d_i$. All bond indices and the physical index are in fact a triple index. The bond dimension can be reduced by applying projectors on the non zero bond indices.
}
\label{fig:tens}
\end{figure}

\section{iPEPS tensors}
\label{app:tensors}
iPEPS tensors, shown in Fig. \ref{fig:tens} are given by the following combination of $F$-symbols and quantum dimensions $d_i$:
\begin{eqnarray}
A^i_{\alpha\beta\gamma} &=& \left(\frac{d_ad_b}{d_c}\right)^{1/4}F^{dab}_{fec}\delta_{aa'}\delta_{bb'}\delta_{cc'}\delta_{dd'}\delta_{ee'}\delta_{ff'} ~~\\
B^i_{\alpha\beta\gamma}  &=& \left(\frac{d_ad_b}{d_c}\right)^{1/4}F^{dab}_{fec}\delta_{aa'}\delta_{bb'}\delta_{cc'}\delta_{dd'}\delta_{ee'}\delta_{ff'} ~~
\end{eqnarray}
By construction each tensor has a triple of bond indices along each of the three bonds towards NN lattice sites. We concatenate each triple into a single bond index, e.g., $\alpha=(a,e,d')$. The physical index is also a triple index $i=(a',b',c')$.
These basic tensors are forming the topological state after proper contraction of bond indices with respect to their triplet structure. For the toric code and double Fibonacci string nets the bond dimension $D=2^3=8$ is redundantly large and can be reduced to $D=4$ and $D=5$ after applying  projectors on the bond indices, namely the only non-zero combinations of bond indices $(i,j,k)$ are those, in which the fusion product $i\times j\times k = 1 +...$ contains the trivial anyon. For the double Ising string net, on the other hand, the original bond dimension $D=3^3=27$ can be reduced to $D=10$.


\section{Algorithm for central idempotents}
\label{app:CI}
\begin{figure}[t]
\centering 
\includegraphics[width = 0.5\columnwidth]{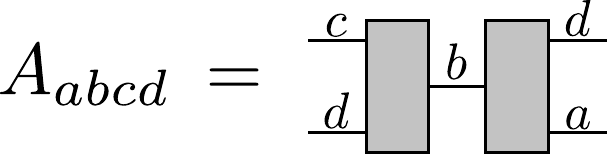}
\caption{The tensors $A_{abcd}$ are proportional to $N^b_{da}N^b_{cd}$ and can be represented by the zippers connected by, but not summed over, the index $b$.}
\label{fig:Atensor}
\end{figure}
In this appendix we present an algorithm for numerical calculation of central idempotents from fusion rules $N^i_{jk}$ and $F$-symbols in a random gauge. When inserted into iPEPS, central idempotents can be thought of as projectors onto minimally entangled states. Central idempotents are build from elements $A_{abcd}$, which form an algebra $\mathcal{A}$ (more precisely it is a C$^*$ algebra). The algorithm can be divided into several points as follows.
\begin{enumerate}

    \item In the first step we determine all non-zero elements of the algebra $\mathcal{A}$ generated by $A_{abcd}$:
    \begin{equation}
        A_{abcd} \propto N^b_{da}N^b_{cd},
    \end{equation}
    as shown in Fig.\ref{fig:Atensor}. Those non-zero elements are the basis vectors of the algebra $\mathcal{A}$, so we can make assigments $e_i= A_{abcd} \neq 0$ and find their multiplication table.
    
    \item When treated as matrices in the $a,c$ indices, the multiplication of $A_{abcd}$ satisfies:
    \begin{equation}
        A_{abcd} A_{efah} = e_i e_j = \sum_k f^k_{ij}~ e_k = \sum_k f^k_{ij}~A_{encl}
    \end{equation}
    Using the tensor network diagrammatic expressions, as shown in Fig.\ref{fig:Aalgebra},  we can derive the formula for the structure factors $f^k_{ij}$, which are given by the $F$-symbols:
    \begin{equation}
        f^k_{ij} = \sqrt{\frac{d_n d_a d_d d_h }{d_f d_b d_l}} F^{dhe}_{nlf} (F^{dah}_{nbf})^{-1} F^{cdh}_{nbl}
    \end{equation}
   
\begin{figure}[t]
\centering 
\includegraphics[width = 0.99\columnwidth]{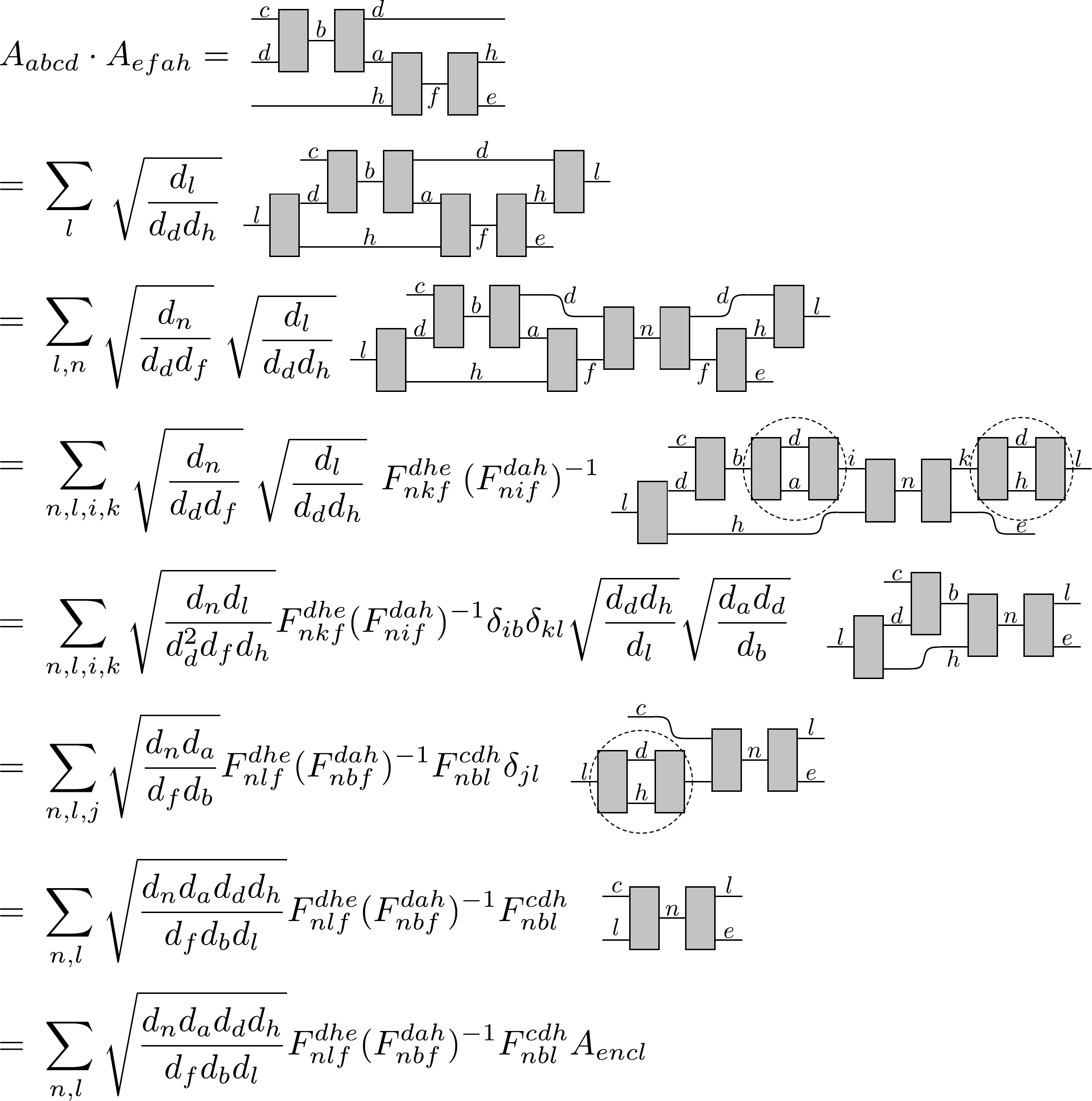}
\caption{Derivation of the algebra generated by the tensors $A_{abcd}$ including proper normalization by the quantum dimensions.}
\label{fig:Aalgebra}
\end{figure}
    
    \item Now we find the center $\mathcal{Z(A)}$ of the algebra $\mathcal{A}$, i.e. we look for such elements $z = \sum_a~z_a e_a$ that $\forall b:~ z e_b = e_b z$. We observe that this simplifies to an equation involving only the structure factors:
    \begin{equation}
        \sum_a~ z_a (e_ae_b - e_be_a) = 0~\Rightarrow~\forall b,c:~ \sum_a~(f^c_{ab} - f^c_{ba})z_a = 0.
    \end{equation}
    Therefore the vector of coefficients of $z$ in $e_a$ basis belongs to the kernel of the matrix $\mathcal{F}$: $c(z):=(z_1,...,z_n) \in \text{Ker}(\mathcal{F})$, whose elements are $\mathcal{F}_{b\oplus c,a} = f^c_{ab}-f^c_{ba}$ with $b\oplus c$ index going through all the combinations of $b,c$ indices.
    
    \item From now on we work only with the commutative algebra $\mathcal{Z(A)}$, with elements $Z_k = \sum_a z^k_a e_a$, which are linear combinations of the original basis with coefficients from $c(z)\in \text{Ker}(\mathcal{F})$. We construct its adjoint representation, which is given by the structure factors
    \begin{equation}
    [ad(Z_k)]_{ab} = \sum_c f^b_{ac} z^k_c
    \end{equation}
    
    \item Due to commutation of the elements of the center $\mathcal{Z(A)}$, if we take random element from the center $Z\in\mathcal{Z(A)}:~ Z=\sum_k c_k ad(Z_k)$ and find the transformation bringing it into the diagonal form: $U^{-1}ZU$, we know that this transformation is diagonalizing all other elements from the center:
    \begin{equation}
        \forall k:~ U^{-1}ad(Z_k)U = D_k,
    \end{equation}
    where $D_k$ is diagonal.
    
    \item We now have to find linear combinations of the matrices $D_k$ to obtain idempotents. Defining $d_k$ as the vectors containing the diagonal elements of $D_k$, finding idempotents boils down to finding orthogonal linear combinations of the vectors $d_k$ that only contain $1$'s and $0$'s. To do this, we build a matrix $D$ with $d_k$ as its row vectors, and compute the row reduced echelon form of the augmented matrix
    \begin{equation}
        D' = [D \, |\, \bbI], \quad \text{rref}(D') = [\text{rref}(D) \, | \, M].
    \end{equation}
    \item The central idempotents are now obtained as
    \begin{equation}
        \mathcal{P}_i = \sum_j M_{ij} Z_j = \sum_{ja} M_{ij} z^j_a e_a.
    \end{equation}
    These central idempotents can be further split into simple idempotents by grouping the different $e_a$ according to the $a=c$ string of the associated tube algebra elements $A_{abcd}$. We note however that this does not always work, as exemplified by the case of $\text{Vec}_{S_3}$ in the main text; an alternative general algorithm will be provided in Ref.\cite{lootens2021topological}.
\end{enumerate}

\end{document}